\documentclass[10pt,journal,compsoc]{IEEEtran}

\pdfoutput=1
\ifCLASSOPTIONcompsoc
  \usepackage[nocompress]{cite}
\else
  \usepackage{cite}
\fi
\ifCLASSINFOpdf
\else
\fi
\usepackage{amsmath}  
  
\newtheorem{definition}{Definition}

\usepackage{algorithm}
\usepackage{algorithmic}

\usepackage{subfigure}
\usepackage{graphicx}
\usepackage{booktabs}
\usepackage{multirow}
\usepackage{enumerate}

\usepackage{amsfonts}

\begin{document}
%
\title{COSINE: Compressive Network Embedding on Large-scale Information Networks}
%
%
%
%

\author{Zhengyan Zhang, Cheng Yang, Zhiyuan Liu, Maosong Sun, Zhichong Fang, Bo Zhang, and Leyu Lin
\IEEEcompsocitemizethanks{\IEEEcompsocthanksitem Zhengyan Zhang, Cheng Yang, Zhichong Fang, Zhiyuan Liu (corresponding author) and Maosong Sun are with the Department of Computer Science and Technology, Tsinghua University, Beijing 100084, China.\protect\\
E-mail: \{zhangzhengyan14, cheng-ya14, fzc14\}@mails.tsinghua.edu.cn, \protect\\liuzy@tsinghua.edu.cn, \protect\\sms@mail.tsinghua.edu.cn
\IEEEcompsocthanksitem Bo Zhang and Leyu Lin are with the Search Product Center, WeChat Search Application Department, Tencent, Beijing 100080, China.\protect\\Email: \{nevinzhang, goshawklin\}@tencent.com}}
\IEEEtitleabstractindextext{%
\begin{abstract}
There is recently a surge in approaches that learn low-dimensional embeddings of nodes in networks. As there are many large-scale real-world networks, it's inefficient for existing approaches to store amounts of parameters in memory and update them edge after edge. With the knowledge that nodes having similar neighborhood will be close to each other in embedding space, we propose COSINE (COmpresSIve NE) algorithm which reduces the memory footprint and accelerates the training process by parameters sharing among similar nodes. COSINE applies graph partitioning algorithms to networks and builds parameter sharing dependency of nodes based on the result of partitioning. With parameters sharing among similar nodes, COSINE injects prior knowledge about higher structural information into training process which makes network embedding more efficient and effective. COSINE can be applied to any \textit{embedding lookup} method and learn high-quality embeddings with limited memory and shorter training time. We conduct experiments of multi-label classification and link prediction, where baselines and our model have the same memory usage. Experimental results show that COSINE gives baselines up to 23\% increase on classification and up to 25\% increase on link prediction. Moreover, time of all representation learning methods using COSINE decreases from 30\% to 70\%.



\end{abstract}

\begin{IEEEkeywords}
Vertex Classification, Link Prediction, Large-scale Real-world Network, Network Embedding, Model Compression
\end{IEEEkeywords}}

\maketitle

\IEEEdisplaynontitleabstractindextext

%
\IEEEpeerreviewmaketitle

\IEEEraisesectionheading{\section{Introduction}\label{sec:introduction}}

%
%
%
%
\IEEEPARstart{T}{here} are various kinds of networks in the real world like computer networks, biological networks and social networks, where elements or users are represented by nodes and the connections between the elements or users are represented by links.
Representing network data is a crucial step before using off-the-shelf machine learning models to conduct advanced analytic tasks such as classification~\cite{tang2015line, perozzi2014deepwalk}, clustering~\cite{chang2015heterogeneous, xie2016unsupervised}, link prediction~\cite{wang2016structural, grover2016node2vec} and personalized recommendation~\cite{barkan2016item2vec,he2017neural}.
Conventional methods present a network by its adjacency matrix, which is hard to be adopted for many machine learning applications due to its sparsity~\cite{perozzi2014deepwalk}. Recently, network embedding, which aims to learn the low-dimensional representation for each vertex in a network, alleviates the sparsity problem and attracts increasing attention.
Network embedding preserves the network structures, the information of nodes~\cite{yang2015network,kipf2016semi} and links~\cite{tu2017transnet} from original networks.
Following the pre-defined proximity measures, similar nodes are mapped to the neighboring regions in the embedding space.

As large-scale online social networks such as Facebook\footnote{http://www.facebook.com}, Twitter\footnote{http://www.twitter.com}, and Sina Weibo\footnote{http://www.weibo.com} are developing rapidly, a large-scale real-world network typically contains millions of nodes and billions of edges.
Most existing network embedding algorithms do not scale for networks of this size. 
There are three reasons: (1) The majority of network embedding algorithms rely on what we call \textit{embedding lookup}~\cite{hamilton2017representation} to build the embedding for each node. We denote the set of nodes by $V$. The mapping function form likes $f(v)$=\textbf{$E\cdot v$}, where $v$ is the target node, $\textbf{E}\in \mathbb{R}^{d\times |V|}$ is a matrix containing the embedding vectors for all nodes, $d$ is the dimension of vectors, $|V|$ is the size of nodes and $\textbf{v} \in \mathbb{I}_{V}$ is a one-hot indicator vector indicating the column of $\textbf{E}$ corresponding to node $v$.
When the size of nodes grows, the dimension of vectors needs to reduce to keep the memory usage not exceed the limit.
On the assumption that we have a network containing 100 million nodes and each node is represented by a 128-dimension floating-point vector, the memory storage of $\textbf{E}$ is more than 100GB.
As the dimension becomes fairly small, the parameters of a model can not preserve enough information about original network and have bad performance on the downstream machine learning tasks.
(2) Most embedding algorithms suffer from the cold-start item problem: if a node has only a few edges to other nodes, chances are that the training of the node's embedding would be insufficient. Broader et al.~\cite{broder2000graph} suggested that the distribution of degrees follows a \textit{power law}, which means there are many low-degree nodes in the large-scale network.
(3) Network embedding on large-scale networks needs to take a long time to train. However, many real-world networks are highly dynamic and evolving over time, so there is a need to speed up training process to follow that.
To sum up, there is a challenge to improve the flexibility and efficiency of large-scale network embedding.

\begin{figure*}[t]
\centering
\subfigure[]{                  
\begin{minipage}{0.4\linewidth}
\centering
\includegraphics[scale=0.2]{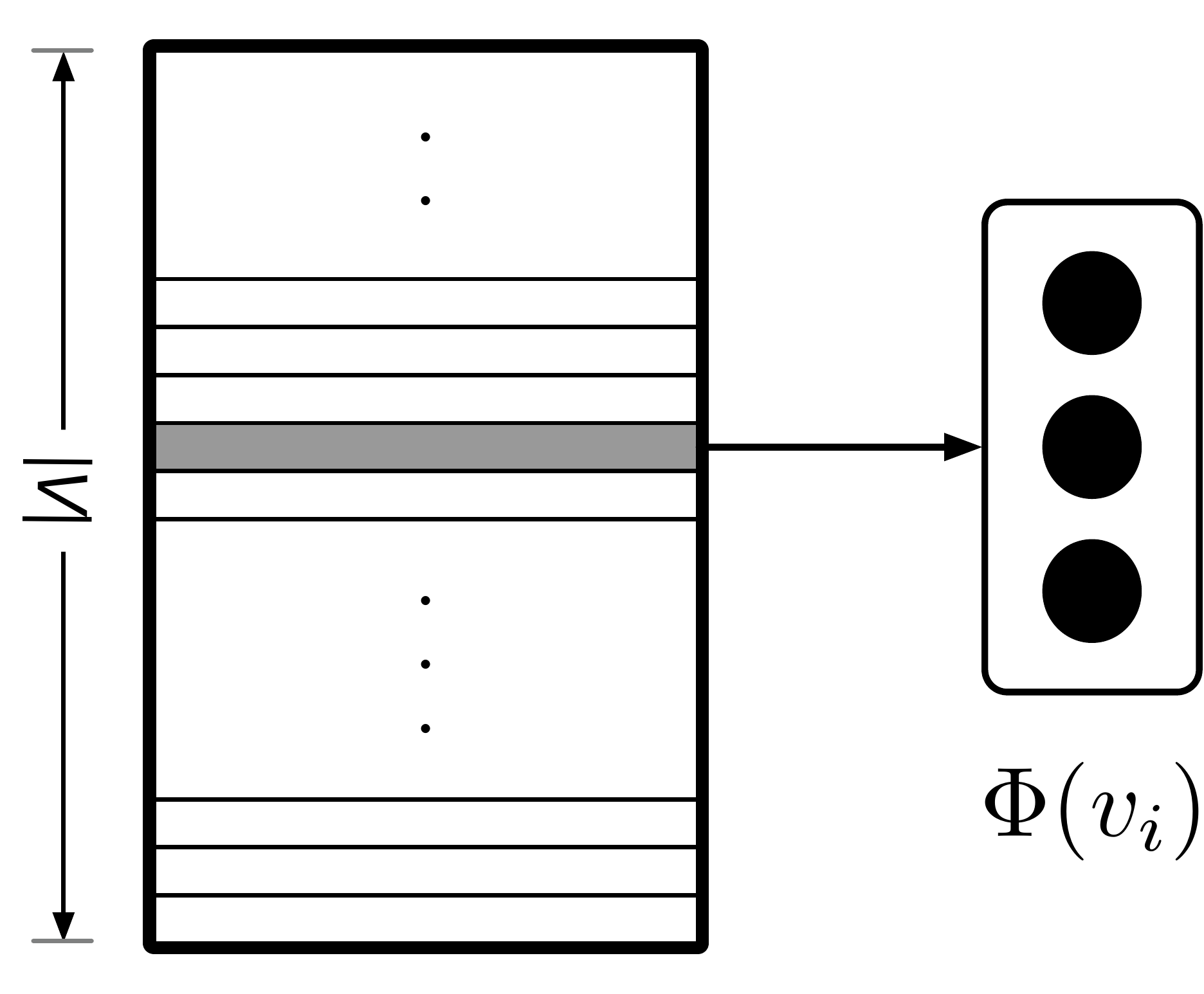}
\label{fig:2:a}         
\end{minipage}}
\subfigure[]{                    
\begin{minipage}{0.5\linewidth}
\centering
\includegraphics[scale=0.2]{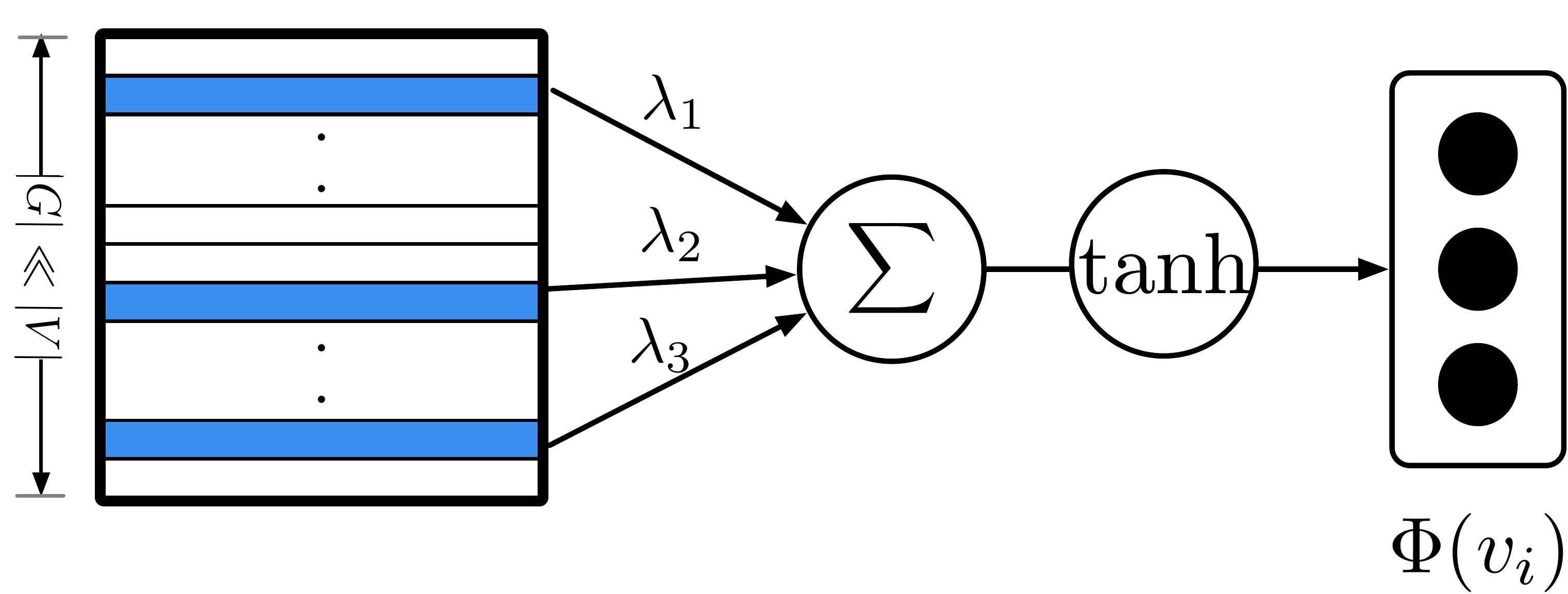}
\label{fig:2:b}             
\end{minipage}}
\caption{Comparison of embedding computations between the conventional approach \ref{fig:2:a} and COSINE approach \ref{fig:2:b} for constructing embedding vectors, where $|\mathcal{G}| \ll |V|$.}                  
\label{fig:2}                                                     
\end{figure*}

In this paper, we explore how to share embedding parameters between nodes, which can address the computational inefficiency of \textit{embedding lookup} methods and be a powerful form of regularization~\cite{hamilton2017representation}.
We assume, in the network, there are groups that contain nodes whose embeddings are partially close to each other and nodes' groups can preserve the information of nodes effectively.
Inspired by this, we propose a unified framework COSINE for compressive network embedding, which improves the quality of embeddings with limited memory and accelerates network embedding process.

It is worth pointing out that COSINE can address all three problems of scalability above.
Firstly, parameters sharing can increase the dimension of vectors without extra memory usage.
The dimension is very critical for preserving similarity from original networks.
Secondly, one input training edge/pair can be used to update two nodes' parameters in previous methods while one edge/pair can be used to update several groups' parameters in COSINE, which affects more than two nodes.
The low-degree nodes also have sufficient training, which solves the cold-start item problem.
Thirdly, the regularization of parameters sharing can be treated as prior knowledge about network structure, which reduces the number of training samples needed.
Since there is a linear relation between running time and training samples, the training time will decrease with COSINE.

We apply COSINE to three state-of-the-art network embedding algorithms, DeepWalk~\cite{perozzi2014deepwalk}, LINE~\cite{tang2015line} and node2vec~\cite{grover2016node2vec}, and conduct experiments on three large-scale real-world networks using the tasks of vertex classification and link prediction where baselines and our model have the same memory usage.
Experimental results show that COSINE significantly improves the performances of three methods, by up to 25\% of AUC in link prediction and up to 23\% of micro-f1 in multi-label classification. Besides, COSINE greatly reduces the running time of these methods(30 \% to 70\% decrease).

To summarize, our major contributions are as follows:

\begin{enumerate}[(1)]
\item We propose a general compressive network embedding framework COSINE, which can work with most existing network embedding algorithms. COSINE compressed the network embedding models by parameters sharing. The parameters sharing is guided by graph  partitioning, which captures the high-order structural information before training.
\item With COSINE, existing methods can learn high-quality embeddings from large-scale network quickly with little memory usage.
\item We conduct two typical network analysis tasks, vertex classification and link prediction. The experimental results show that COSINE achieves significant and consistent improvements on state-of-the-art methods.
\end{enumerate}

\section{Related Work}
We provide a framework of network representation learning which leverages the similarities between the nodes.
The latent low-dimensional representation of nodes in networks could be used in network analysis applications, such as link prediction, vertex classification, graph partitioning and so on.
In what follows, we will introduce a brief overview of related work in the aforementioned tasks and some methods.

\subsection{Link Prediction}
Link prediction aims to predict the likelihood of the existence of edges between the nodes, so it is often modeled as a social recommendation problem.
Intuitively, the nodes with higher affinities would have much greater chances to be connected than those with lower affinities.

The problem was first formally introduced for mining social networks in~\cite{liben2007link}. 
Traditionally, there are some topology-based methods. 
Zhou et al.~\cite{zhou2009predicting} leverage the resource allocation index in networks.
However, due to the lack of topological information, Hasan et al.~\cite{al2006link} extract contextual information from the nodes and edges.
Liang et al.~\cite{liang2017link} combined topological information with attributes of nodes to improve the accuracy significantly. 
Matrix and tensor factorizations methods~\cite{menon2011link, dunlavy2011temporal} have also been applied to evaluate the likelihood.
Besides, meta-paths~\cite{sun2013mining} are critical elements of link-prediction models in heterogeneous networks.

\subsection{Vertex Classification}
Vertex classification is one of the most common semi-supervised tasks in network analysis, which aims to classify the vertices to at least one groups.
The application of the task could be shown in many areas, such as protein classification~\cite{grover2016node2vec}, user profiling~\cite{tang2010combination, li2014user}, and so on.

There are several traditional approaches to address the classification problem, such as iterative methods~\cite{sen2008collective,zhu2002learning}. Further, hand-crafted features are usually leveraged~\cite{Chittaranjan:2011un, schwartz2013personality}, but the methods are unadaptable to multiple real-world scenarios.

Recent years, node embedding has been introduced to solve the problem~\cite{perozzi2014deepwalk, tang2015line, grover2016node2vec}.
For instance, the learned low-dimensional representations could be inputted to traditional classifier such SVM.
The main advantages of the method are the high computation efficiency and the robustness when encountered with data sparsity.
Besides, Yang et al.~\cite{yang2016revisiting} jointly learn the node embedding and train the classifier to enhance the performance of vertex classification.

\subsection{Network Representation Learning}
The goal of Network Representation Learning(NRL) is to map the nodes in networks into low-dimensional vector space while trying to preserve the properties of networks.
NRL has been widely used by machine learning models due to its effectiveness and ease of use.

Most current node embedding techniques are lookup algorithms, i.e., there is a matrix containing the embedding vectors for all nodes, so we just need to look up in the matrix for a specific embedding. 
Early works in NRL mainly are based on the factorization of the graph Laplacian matrix, such as Isomap~\cite{tenenbaum2000global}, Laplacian Eigenmaps~\cite{belkin2002laplacian} and Social Dimension~\cite{tang2009relational}.
However, the computational expense of those approaches is so high that they could not be adapted to large-scale networks.
Inspired by word embedding methods in Natural Language Processing, DeepWalk~\cite{perozzi2014deepwalk} and node2vec~\cite{grover2016node2vec} combine word2vec~\cite{mikolov2013distributed} with different random walk strategies. 
Tang et al.~\cite{tang2015line} design a model called LINE, which leverages first- and second-order proximities between two vertices.
Furthermore, Wang et al.~\cite{wang2016structural} extend the aforementioned model with neuron networks to learn non-linear features.
Besides the look-up algorithms, Graph Convolutional Networks (GCN)~\cite{kipf2016semi} and GraphSAGE~\cite{hamilton2017inductive} are paradigms of neighborhood aggregation algorithms. 
They generate node embeddings with information aggregated from a node’s local neighborhood and some shared parameters.

Closely related to our model, HARP~\cite{chen2017harp} first coarsens the graph, and after that, the new graph consists of supernodes. 
Afterward, network embedding methods are applied to learn the representations of supernodes, and then with the learned representation as the initial value of the supernodes' constituent nodes, the embedding methods are run over finer-grained subgraphs again.
Compared with HARP, MILE~\cite{liang2018mile} implements embeddings refinement to learn better representations for nodes in finer-grained networks with lower computational cost and higher flexibility.
While HARP and MILE still follow the setting of embedding lookup as previous work did, our framework manages to reduce the memory usage as well as improve the scalability.

\subsection{Graph Partitioning}
Graph partitioning aims to partition the vertices in a network into \textit{k} disjoint sets so that unbalances of the sets' size and the total weight of edges cut by the partition are minimized.
Well-known techniques in graph portioning include Metis~\cite{karypis1998fast}, Scotch~\cite{chevalier2008pt} and KaHIP~\cite{sanders2011engineering}. 

One of the fastest available distributed memory parallel code is ParMetis~\cite{karypis1999parallel}, the parallel version of Metis, but it is unable to maintain the balance of the blocks.
Besides, the greedy local search algorithms adopted by ParMetis are so simple that the performance of the partition remains unsatisfied.
LaSalle et al.~\cite{lasalle2013multi} propose mt-metis, which is a shared-memory parallel version of the ParMetis.
Avoiding message passing overheads and modifying the algorithms used in ParMetis enable mt-metis to run faster and demand less memory.
KaFFPa~\cite{sanders2011engineering} is a framework of multilevel graph partitioning equipped with new local improvement methods ad global search strategies transferred from multi-grid linear solvers.
The framework also runs fast and improves the results of other petitioners.
Recently, Meyerhenke et al. present ParHIP~\cite{meyerhenke2017parallel}， which can be applied to large complex networks.
Their strategies to overcome complex networks are parallelizing and adapting the label propagation technique.
The quality obtained by ParHIP is higher than aforementioned state-of-the-art systems such ParMetis.

\subsection{Model Compression}
Model Compression focuses on building a light-weight approximation of the original model, whose size is reduced while preserving accuracy.

Compression for Convolutional Neural Networks (CNN) has been extensively studied, mainly divided into three following branches. 
First, Low-rank matrix/tensor factorization~\cite{sainath2013low,jaderberg2014speeding,denil2013predicting} is derived on the assumption that using a low-rank approximation of the matrix to approximate each of the networks' weight matrices. 
Second, network pruning~\cite{han2015deep, han2015learning, see2016compression, zhang2017towards} removes trivial weights in the neural network to make the network sparse.
Third, network quantization reduces the number of bits required to represent each weight, such as HashedNet~\cite{chen2015compressing} and QNN~\cite{hubara2016quantized}.
There are also several techniques to compress word embeddings.
Character-based neural language models~\cite{kim2016character, botha2017natural} reduce the number of unique word types, but are faced with the problem that Eastern Asian languages such as Chinese and Japanese have a large vocabulary.
Kept out of the problem,~\cite{shu2017compressing} adopts various methods involving pruning and deep compositional coding to construct the embeddings with few basis vectors. 
Besides, Word2Bits~\cite{lam2018word2bits} extends word2vec~\cite{mikolov2013distributed} with a quantization function, showing that training with the function acts as a regularizer.

\section{Problem Definition}

In this section, we introduce some background knowledge and formally define the problem of compressive network embedding. We formalize the network embedding problem as below:

\begin{definition}
\textbf{(Network Embedding)} Given a network $G=(V,E)$, where $V$ is the set of nodes and $E$ is the set of edges, the goal of \textbf{network embedding} is to learn a mapping function $\Phi : V \mapsto \mathbb{R}^{|V|\times d}$, $d \ll |V|$. This mapping $\Phi$ defines the embedding of each node $v \in V$.
\end{definition}

The parameter of the mapping function $\Phi$ in most existing approaches is a matrix which contains $|V|\times d$ elements, as they embed each node independently. However, there are many large networks which contain billions of nodes and their scales are still growing. It is difficult to store all embeddings in the memory when we train a large network embedding model. In this work, we hypothesize that learning embeddings independently causes redundancy in the parameter, as the inter-similarity among nodes is ignored. For example, Jiawei Han and Philip S. Yu both belong to the Data Mining community in an academic network so that the embedding of Data Mining community can be the shared part of their embeddings. Therefore, we propose the problem of compressive network embedding to reduce the memory use by sharing parameters between similar nodes and make large network embedding available.

There are two main challenges of \textit{compressive network embedding} : (1) how to find the shared and independent parts of embeddings between similar nodes, (2) how to combine different parts of embeddings for each node to generate the node representation. For example, though Jiawei Han and Philip S. Yu may both have the embedding of Data Mining community in their shared part, the other portion in both vectors still has to be trained independently to capture extra information about their academic life. Following the intuition of creating partially shared embeddings, we represent each node $v$ with a group set $S_v=(\mathcal{G}_1, \mathcal{G}_2, \mathcal{G}_3,\ldots, \mathcal{G}_M)$. Each group is a social unit where each node has a strong relation to each other and we denote the set of all groups by $\mathcal{G}$. We hypothesize that $M$ groups are enough to figure out the characteristic of each node in networks. Having the same group set doesn't mean they have the same embedding, as different people have different preferences among groups. Therefore, the model also needs to learn the preference of each node while training. In summary, we formally define the \textit{compressive network embedding} problem as follows:

\begin{definition}
\textbf{(Compressive Network Embedding)} Given a network $G=(V,E)$ and the dimension of embedding $d$, the goal of \textbf{compressive network embedding} is to learn a network embedding model which has less than $d|V|$ parameters (the traditional look-up methods need a memory space for $|V|$ $d$-dimension vectors). At the same time, the learned model can represent each node by a $d$-dimension vectors.
\end{definition}

\section{Method}

\begin{figure*}[t]
\centering
\subfigure[Partitioning a simple graph.]{                  
\begin{minipage}{0.3\linewidth}
\centering
\includegraphics[scale=0.3]{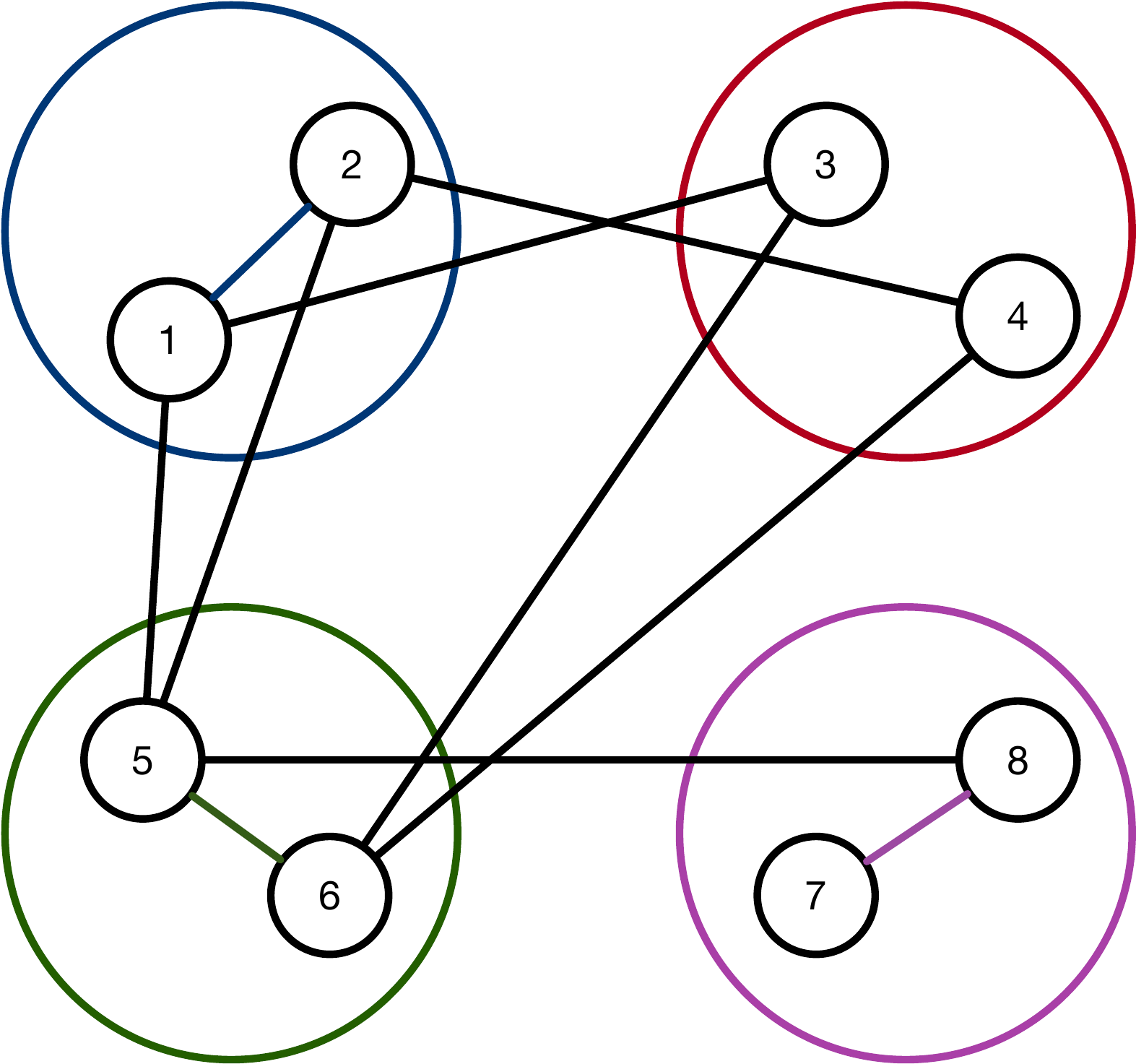}
\label{fig:1:a}         
\end{minipage}}
\subfigure[Building group set by random walk.]{                    
\begin{minipage}{0.3\linewidth}
\centering
\includegraphics[scale=0.3]{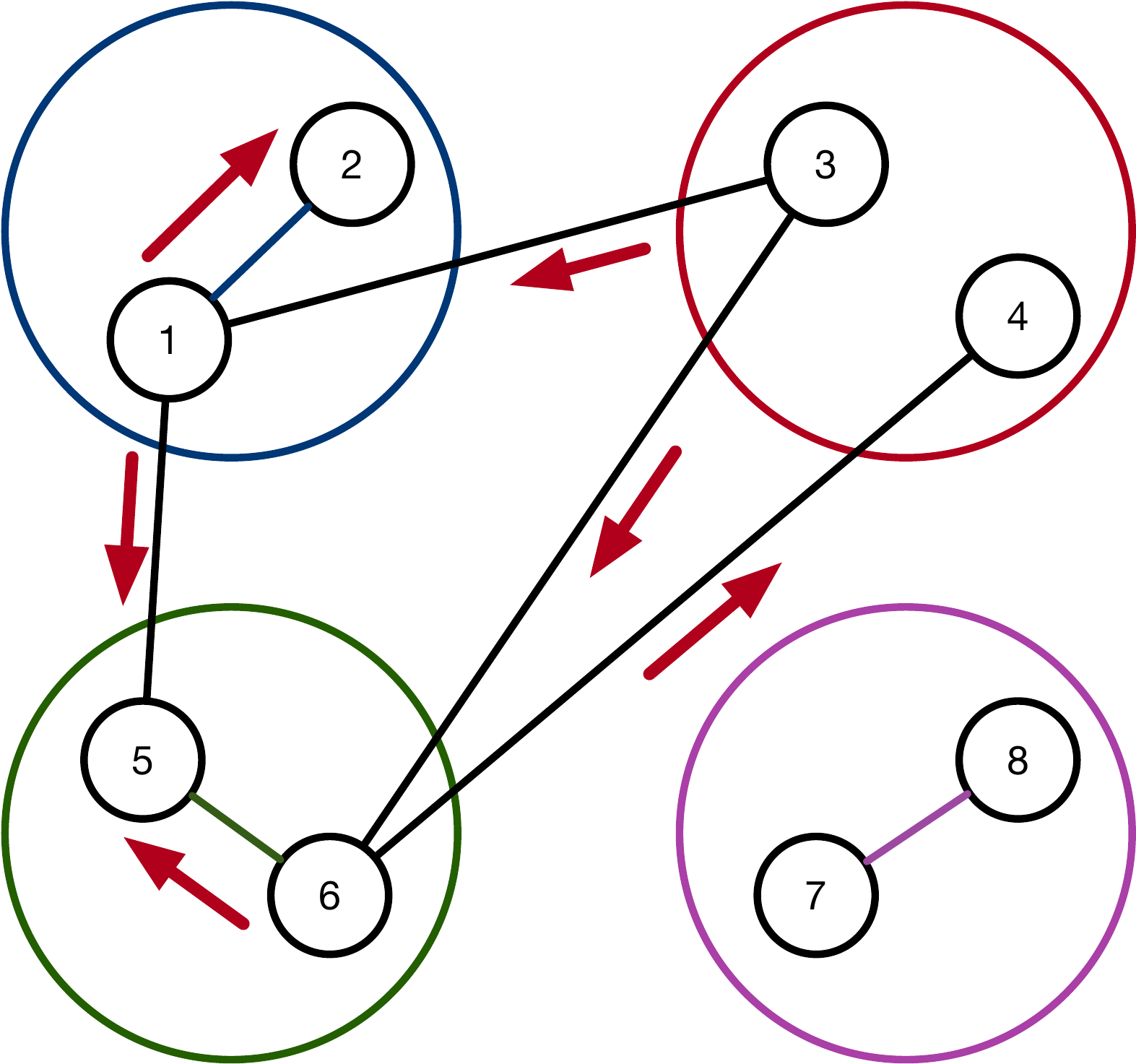}
\label{fig:1:b}             
\end{minipage}}
\subfigure[Group set of node 3.]{                   
\begin{minipage}{0.3\linewidth}
\centering                                  
\includegraphics[scale=0.3]{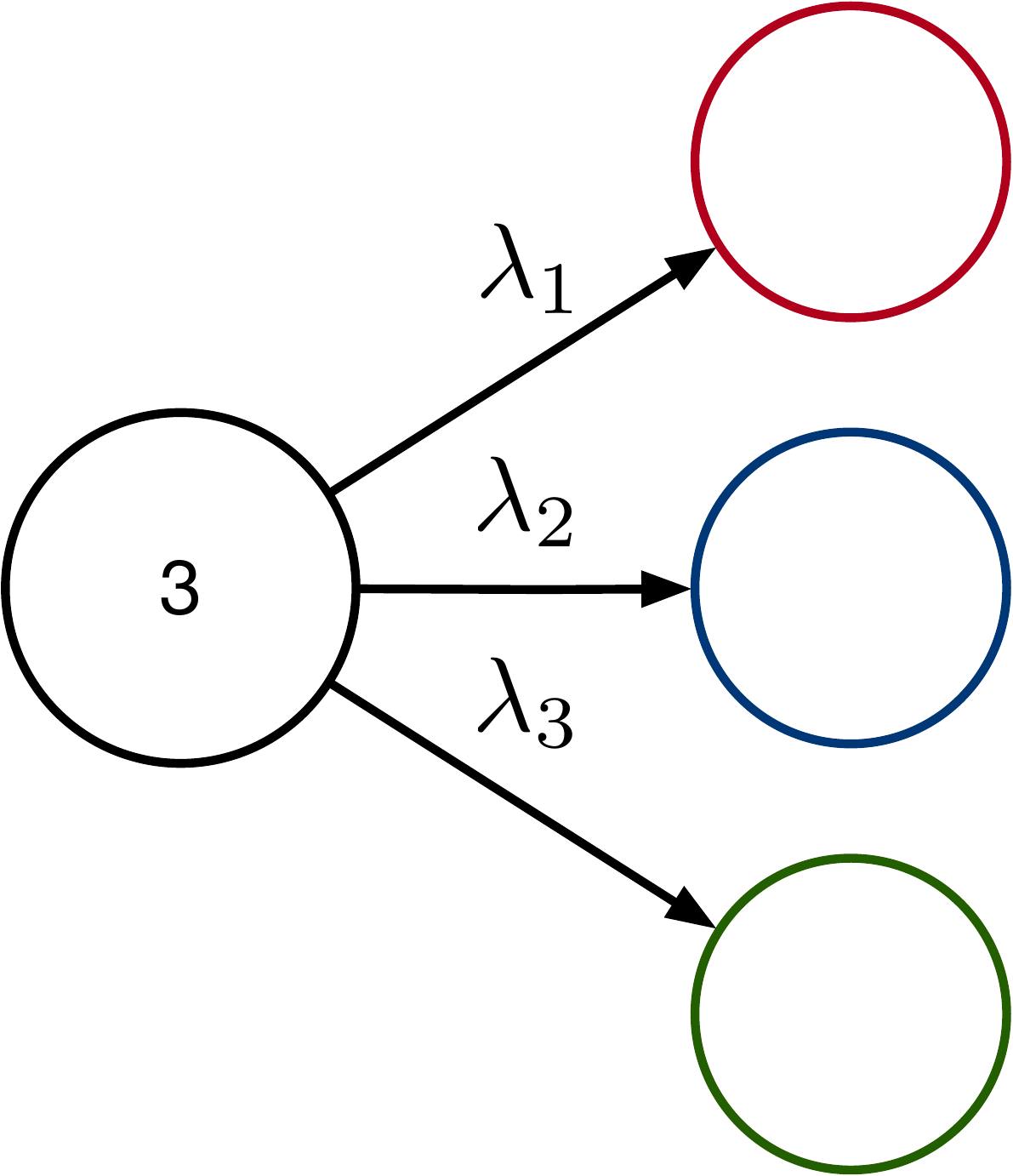} 
\label{fig:1:c}      
\end{minipage}}
\caption{A toy example of graph partitioning and group mapping. After partitioning, there are four groups in \ref{fig:1:a} which are presented in different colors. We begin several random walks rooted at Vertex 3 to sample its group set and set walk length as 2 in \ref{fig:1:b}. According to the result of random walks, we find three groups which are related to Vertex 3 and use these three groups to present Vertex 3. Vertex 3 will share its embeddings parameters with vertices whose group sets also contain some of these three groups.}                  
\label{fig:1}                                                     
\end{figure*}

In this section, we present a general framework which can cover several Network Embedding algorithms including LINE, DeepWalk and Node2Vec and learn better embeddings with limited memory resources. Our general framework consists of the following steps: (1) use Graph Partitioning methods to find vertices' partition/group from the real network; (2) for each vertex, sample intelligently a set of groups to build the group mapping function; (3) for each vertex, use an architecture based on Graph Convolutional Network (GCN)~\cite{kipf2016semi} to aggregate information from its group set and output an embedding for each vertex; (4) use different network embedding objective functions to train the model (take SGNS as example). We explain each stage in more detail below.

\subsection{Graph Partitioning}

Graph partitioning is used to divide a network into several partitions/groups, which is important for parameter sharing and incorporating high order structure before training. There are explicit groups in social networks which consist of people who share similar characteristics and collectively have a sense of unity. As we don't have the information about existing social groups, we should use graph partitioning methods to assign each node one group. Based on the original groups, COSINE can sample more related group from network structure.

There are two kinds of methods which are able to assign groups to nodes: (1) overlapping methods like AGM~\cite{yang2012community}, where a node can belong to multiple groups at once; (2) non-overlapping methods like Graph Coarsening and Graph Partitioning, where a node only belongs to one group. We present each node with at most $M$ different groups while overlapping methods can't limit the quantity of groups for each node. So, we choose non-overlapping methods in our framework.

HARP~\cite{chen2017harp} and MILE~\cite{liang2018mile} have used Graph Coarsening to find a smaller network which approximates the global structure of its input and learn coarse embeddings from the small network, which serve as good initializations for learning representation in the input network. Graph Coarsening coarsens a network without counting the number of origin nodes that belong to a coarsen group. It leads to the unbalance among coarse groups. For instance, there may be some groups which only have one node, which results in defective parameters sharing.

In our framework, we use Graph Partitioning method to assign each node a specific group. Graph Partitioning~\cite{sanders2011engineering} is mostly used in \textit{high performance computing} to partition the underlying graph model of computation and communication. Graph Partitioning divides nodes into several groups and encourages more edges in groups and fewer edges between groups. Each node links to the nodes in its group strongly and links to the rest nodes weekly. The advantage of graph partitioning is that it divides the graph vertex set into $k$ disjoint groups of a roughly \textbf{equal size} which benefits the parameters sharing.

\subsection{Group Mapping}

After graph partitioning, we have a mapping $g(v)$ from one node to one group. In our framework, we plan to use a group set to present a node instead of just a group. For each node, $g(v)$ plays an important role in the construction of group set $S_v$. However, We also need to find more related groups for each node.

We hypothesize that the neighbors' groups are characteristics for a node. There is proximity between nodes and their neighbors so that the neighbors' groups are also useful to present nodes. To introduce high order proximity into our model, we consider not only one-hop neighbors but also k-hop neighbors. We denote the $j$th random walk rooted at vertex $v_i$ as $W_{v_i}^j$. It is a stochastic process with random variables $v^1, v^2, \ldots, v^k$ such that $W_{v_i}^k$ is a k-hop neighbor for the root vertex $v_i$. There are two advantages of random walks for find neighbors' groups. First, random walks have been used to extract local structure of a network~\cite{perozzi2014deepwalk} and achieve big success. Different from Broad First Search (BFS), the walker can revisit a vertex in a walk which means this vertex is important for the local structure and we should pay more attention to it. Second, several walkers can walk simultaneously on a network. As we plan to solve the problem of compressive network embedding on large-scale networks, it is important for group mapping to run parallelly.

We denote the number of groups in a node group set as $|S_v|$. After several walks rooted at $v_i$, we have a node set which contains the neighbors of $v_i$ in k-hop. By the mapping $g(v)$, we have a group set $S_{raw}$ which contains the neighbors' groups. In practice, the size of $S_{raw}$ is usually bigger than $|S_v|$. So, we have to select the $|S_v|$ most related groups from $S_{raw}$ according to the groups' frequency in walks. Algorithm \ref{alg1} presents our group mapping algorithm. The function \textit{Concatenate} is used to join two walk lists and the elements in $W_{v_i}$ increased by walk length $k$. The function \textit{SelectByFrequency} chooses the top $n$ frequent items in the input list. 

Fig. \ref{fig:1} presents an illustrative example for graph partitioning and group mapping. Vertex 3 is far away from the pink group, as no walker can arrive at Vertex 7 and 8 in two hops. And there exist communications between the rest groups and Vertex 3 on the graph which is consistent with the mapping result. So, group mapping can introduce high proximity of network structure to the mapping function $\Phi_v$ and motivate the partial embedding sharing. $\lambda_i^v$ presents Vertex 3's preference for a specific group which is unknown to the mapping function and need to be learned.

\begin{algorithm}
\caption{Group Mapping($G, \Phi_V$)}
\label{alg1}  
\begin{algorithmic}[1]
\REQUIRE  \hspace*{0.02in} \\ 
graph $G(V, E)$ \\
one-one mapping from nodes to groups $g(v)$ \\
walk per vertex $\gamma$ \\
walk length $k$ \\
the number of groups in each set $n$ \\
\ENSURE mapping from nodes to group sets $\Phi_V$
\STATE Initialize $\Phi_V$
\FOR{\textbf{each }$ v_i \in V$}
\STATE $W_{v_i} \gets \{\}$ 
\FOR{$j=0$ to $\gamma$}
\STATE $W_{v_i}^j \gets RandomWalk(G, v_i, t)$
\STATE $W_{v_i} \gets Concatenate(W_{v_i}, W_{v_i}^j)$ 
\ENDFOR
\STATE $\mathcal{G}_{v_i} \gets g(W_{v_i})$
\STATE $S_v \gets SelectByFrequency(\mathcal{G}_{v_i}, n)$
\STATE $\Phi_V(v_i) \gets S_v$
\ENDFOR
\end{algorithmic}  
\end{algorithm}  

\subsection{Group Aggregation}

Graph aggregation is the process where the model takes the group set as input and outputs the node embedding. We have a fixed number of groups for each node, which is convenient for us to aggregate groups' embeddings. However, we need a method to aggregate groups' embeddings into a single representation that remain invariant to the order of groups, as there is no sense of groups' ordinality. In practice, there are various ways to aggregate embeddings that meet the requirement. Common methods include: (1) use an RNN to encode the group embedding one-by-one as a sequence, and augment with  different permutations of the features so that ideally the RNN learns an order-invariant model for group aggregation; (2) use a symmetric function $f(g_1, g_2, \ldots, g_n)$, whose value does not change when the input $\{g_i|1\le i \le n\}$ is permuted. We now discuss these methods in detail.

To train an order-invariant RNN model, we have to augment the training data by adding many random permutations. It leads to high computation overhead which is unacceptable for the large network. Our framework is designed to run fast and memory-free network embedding algorithm in a big network dataset.

In computer vision, using max-pooling as a feature aggregator has shown good performance~\cite{qi2017pointnet}. However, max-pooling tends to select the positive features and ignore the negative features among groups. The aggregated representation by max-pooling has more positive entries than negative entries so that it's not for the downstream tasks.

Our framework aggregate groups' feature via computing a weighted average of embeddings to estimate the optimal aggregation function. This method has been used in the Graph Convolutional network (GCN)~\cite{kipf2016semi} and GraphSAGE~\cite{hamilton2017inductive}. For each node, it has an aggregation kernel whose size is equal to the number of groups $|S_v|$. We denote the kernel as $K_v=(\lambda_1^v, \lambda_2^v, \ldots, \lambda_{|S_v|}^v)$ where $\lambda$ is a scalar learned from the network. To aggregate groups, we use the following equation:
\begin{equation}
  f(g_1, g_2, \ldots, g_{|S_v|}) = \sum_{i=1}^{|S_v|}\lambda_i^v \Phi_\mathcal{G}(g_i)
\end{equation}
where $\Phi_\mathcal{G}$ denotes the embedding mapping of groups, and there is no regularization on the sum of $\lambda_i^v$. To prevent the gradient explosion in the early stages of training, we use $\tanh$ as the activation function to regularize the entry value of $f(g_1, g_2, \ldots, g_{|S_v|})$. We have the final group aggregation function as follow:
\begin{equation}
  f(S_v) = \tanh(\sum_{i=1}^{|S_v|}\lambda_i^v \Phi_\mathcal{G}(g_i)), \quad g_i \in S_v
\end{equation}

\subsection{Objective Function and Optimization}

COSINE can apply to most existing network embedding algorithms by using the aggregated embeddings instead of the look-up embeddings. In this subsection, we take skip-gram objective function with negative sampling (SGNS) as an example to illustrate how COSINE tunes parameters of group embeddings and aggregation function via stochastic gradient descent. SGNS is the most common graph-based loss function:
\begin{equation}
\begin{aligned}
  \mathcal{L}(u, v) = &-log(\sigma(f_C(S_u)^\top f(S_v))) \\
  &- \sum_{i=1}^K E_{v_n \sim P_n(v)} [log(\sigma( f_C(S_{v_n})^\top f(S_v) ))]
\end{aligned}
\label{loss}
\end{equation}
where $u$ is a node that co-occurs near $v$ ($u$ and $v$ co-occur at the same window in random-walk based methods and $u$ is $v$'s neighbor in LINE. ), $\sigma$ is the sigmoid function, $P_n$ is a negative sampling distribution, $f_C$ is the aggregation function for context embedding, and $K$ defines the number of negative samples. Importantly, unlike embedding look-up approaches, we not only share parameters among similar nodes via group embedding but also use the same aggregation kernel while combining vertex embedding and context embedding~\cite{mikolov2013distributed}.

We adopt the asynchronous stochastic gradient algorithm (ASGD)~\cite{recht2011hogwild} which is broadly used in learning embeddings for optimizing Equation\ref{loss}. In each step, we sample an edge $(v_i, v_j)$ as a mini-batch and the gradient w.r.t the group embedding $e_g$ which belongs to one of the vertex $v_j$'s groups will be calculated as:
\begin{equation}
\begin{aligned}
\frac {\partial O} {\partial e_g} = &-[1-\sigma(f_C(S_{v_i})^\top f(S_{v_j}))] \\
&f_C(S_{v_i}) \otimes (1-f(S_{v_j}) \otimes f(S_{v_j})) * \lambda_g^{v_j}
\end{aligned}
\end{equation}
where $\otimes$ is element-wise multiplication, and $\lambda_g$ is the kernel parameter for group embedding $e_g$. To conduct a stable update on group embedding, we will adjust the learning rate according to the Euclidean norm of aggregation kernel. If the kernel norm is big, we adjust the learning rate smaller to make the shared parameter change smoothly.

Besides computing the gradient of group embedding, we also need to compute the gradient of aggregation kernels for nodes:
\begin{equation}
\begin{aligned}
\frac {\partial O} {\partial \lambda_g^{v_j}} = &-[1-\sigma(f_C(S_{v_i})^\top f(S_{v_j}))] \\
&\sum [f_C(S_{v_i}) \otimes (1-f(S_{v_j}) \otimes f(S_{v_j})) \otimes e_g]
\end{aligned}
\end{equation}
We find if the group embedding $e_g$ is similar to the node $v_i$'s embedding, the aggregation kernel tend to put more weight on this group. After updating aggregation kernel, the model can learn the node's preference among groups. However, the gradient will explode when $e_g$ and $f_C(S_{v_i})$ are nearly in the same direction. Instead of using gradient clipping, we consider the kernel gradient globally. If all gradients are large, we should adjust the learning rate smaller, and if all gradients are small, we should find the more important group and update its kernel with a higher learning rate.

\section{Experiments}

We empirically evaluated the effectiveness and efficiency of COSINE. We applied the framework to three \textit{embedding lookup} methods. According to the experimental results on three large-scale social network, our framework can improve the quality of embeddings with the same memory usage and reduce the running time.

To evaluate the quality of embeddings, we conduct two kinds of network analysis tasks, multi-label classification and link prediction. We treat nodes' embeddings as their features in the downstream machine learning tasks. The more beneficial the features are to the tasks, the better quality the embeddings have.

\subsection{Datasets}

An overview of the networks we consider in our experiments is given in Table \ref{tab:network}.

\textbf{Youtube~\cite{tang2009relational}} contains $1,138,499$ users and $4,945,382$ social relations between them, which is crawled from the popular video sharing website. The labels represent groups of users that enjoy common video categories.

\textbf{Flickr~\cite{tang2009relational}} contains $1,715,255$ users and $22,613,981$ social relations between them, which is crawled from the photo sharing website. The labels represent the interest groups of users such as \textit{"black and white photos"}.

\textbf{Yelp~\cite{liang2018mile}} contains $8,981,389$ users and $39,846,890$ social relations between them. The labels here represent the business categories on which the users have reviewed.

\begin{table}[!htb]
\centering
\caption{Networks used in our experiments.}
\begin{tabular}{l|l|l|l}
Name   & Youtube & Flickr & Yelp\\ 
\hline
$|V|$ & $1,138,499$ & $1,715,255$ &$8,981,389$\\
$E$ & $4,945,382$ & $22,613,981$&$39,846,890$\\
\#Labels &$47$ & $20$&$22$\\
Directed & directed &directed&undirected\\
\end{tabular}
\label{tab:network}
\end{table}

\subsection{Baselines and Experimental Settings}

To demonstrate that COSINE can work with different graph embedding methods, we explore three popular state-of-the-art methods for network embedding.

\textbf{LINE~\cite{tang2015line}} learns two separate network representations LINE$_{1st}$ and LINE$_{2nd}$ respectively.LINE$_{1st}$ can be only used on undirected networks and LINE$_{2nd}$ is suitable for undirected and directed networks. We choose LINE$_{2nd}$ as the baseline in our experiments.

\textbf{DeepWalk~\cite{perozzi2014deepwalk}} learns network embeddings from random walks. For each vertex, truncated random walks starting from the vertex are used to obtain the contextual information.

\textbf{Node2vec~\cite{grover2016node2vec}} is an improved version of DeepWalk, where it generates random walks with more flexibility controlled through parameters $p$ and $q$. We use the same setting as DeepWalk for those common hyper-parameters while employing a grid search over return parameter and in-out parameter $p, q \in \{0.25, 0.5, 1, 2, 4\}$.

\textbf{Experimental Settings} DeepWalk uses hierarchical sampling to approximate the softmax probabilities while hierarchical softmax is inefficient when compared with negative sampling~\cite{mikolov2013distributed}. To embed large-scale network by DeepWalk, we switch to negative sampling, which is also used in node2vec and LINE, as ~\cite{grover2016node2vec} did. We set window size $w=5$, random walk length $t = 40$ and walks per vertex $\gamma = 5$ for random-walk based methods.
And, we use default settings~\cite{tang2015line} for all hyper-parameters except the number of total training samples for LINE. 
We found these settings are effective and efficient for large-scale network embedding. With COSINE, network representation learning model can use less memory to learn the same dimension embeddings. For instance, if we set the dimension of embeddings as $d$, we need $2d|V|$ floating-point numbers to store the uncompressed model and $|S_v||V| + 2d|\mathcal{G}| \approx |S_v||V|$ for compressed model, where $|\mathcal{G}| \ll |V|$. And $|S_v|$ is also a small value, which means compressed model takes $\frac {|S_v|} {2d}$ times less space than uncompressed model. As we focus on the problem of large network representation learning with limited memory and evaluate uncompressed and compressed models fairly, we use different dimension in compressed and uncompressed model to make sure the memory usage is the same. We set $d=100$ for uncompressed models and $d=8$ for compressed models and adjust the number of groups for each dataset to keep them the same memory. Note that the size of the group set for each node $|S_v|$ is $5$ for all dataset, which we assumed enough for representing nodes' structural information.

In previous works~\cite{tang2015line,grover2016node2vec}, they generate an equal number of samples for each method which means some methods may not train to convergence. We plan to prove our framework can improve the capacity of each method. So, different methods should have different numbers of samples to guarantee that they can train to convergence. For \textit{LINE}, we define one epoch means all edges have been trained one time. For \textit{DeepWalk} and \textit{node2vec}, we regard traversing all walks for training as an iteration. We train each model with different epochs or iterations to find the best number of samples. For random walk based models with \textit{COSINE} framework, we found the best number of samples is one iteration on part of the random walks. In other words, there is no need to do a complete iteration. For instance, $0.2$ iteration means that the model just takes one iteration on $20\%$ walks. As a framework, compressed and uncompressed model use the same random walks to ensure that the input data is the same.

\subsection{Link Prediction}

Link prediction task can also show the quality of network embeddings. Given a network, we randomly remove $10\%$ links as the test set and the rest as the training set. We treat the training set as a new network which is the input of network representation learning and employ the representations to compute the similarity scores between two nodes, which can be further applied to predict potential links between nodes.

We choose three kinds of similarity score functions:
\begin{itemize}
\item L1-norm: $f(e_1, e_2)=\|e_1-e_2\|_1$
\item L2-norm: $f(e_1, e_2)=\|e_1-e_2\|_2$
\item Dot product: $f(e_1, e_2)=e_1^Te_2$
\end{itemize}
where $e_1, e_2$ are the embeddings of two given nodes. For dot product function, the higher score indicates two nodes have more affinities while the lower score indicates two nodes are more similar for L1-norm and L2-norm function.

We employ two standard link prediction metric, AUC~\cite{hanley1982meaning} and MRR~\cite{voorhees1999trec}, to evaluate compressed and uncompressed methods. Given the similarity of all vertex pairs, Area Under Curve (AUC) is the probability that a random unobserved link has higher similarity than a random nonexistent link. Assume that we draw $n$ independent comparisons, the AUC value is
\[
  AUC = \frac {n_1+0.5n_2} {n}
\]
where $n_1$ is the times that unobserved link has a higher score and $n_2$ is the times that they have an equal score. The mean reciprocal rank (MRR) is a statistical measure for evaluating the ranks of the unobserved links' scores among scores of random nonexistent links, where all links have the same head node and different tail nodes. Assume that we sample a set $Q$ of unobserved links, the MRR value is
\[
  MRR = \frac 1 {|Q|}\sum_{i=1}^{|Q|}\frac 1 {rank_i}
\]
where $rank_i$ is the rank of $i$th unobserved link.

We show the AUC values of link prediction on different datasets in Table \ref{table:result-lp-auc} and the MRR values in Table \ref{table:result-lp-mrr}. From these tables, we observe that:

\begin{enumerate}[(1)]
\item The proposed COSINE framework consistently and significantly improves all baseline methods on link prediction. Especially, in Youtube, LINE with COSINE gives us nearly $4\%$ gain over best baseline in AUC score and gives us over $7\%$ gain over best baseline in MRR score. 
\item Dot product is the best score function. For Youtube and Yelp datasets, COSINE-LINE using Dot product is the best method while COSINE-N2V using Dot product is the best one for Flickr. 
\end{enumerate}

In summary, COSINE gives us the best result of AUC and MRR, which means the high order proximity encoded before training is essential for measuring the similarity between nodes precisely.

\begin{table}[!htb]
  \caption{Area Under Curve (AUC) scores for link prediction. Comparison between compressed and uncompressed models using binary operators: (a) L1-norm, (b) L2-norm, and (c) Dot product.} 
  \label{table:result-lp-auc}
  \begin{center}
  \begin{tabular}{c|c|c|c|c}
  \multirow{2}{*}{\textbf{Op}} & \multirow{2}{*}{\textbf{Algorithm}} & \multicolumn{3}{c}{\textbf{Dataset}} \\
  & & Youtube & Flickr & Yelp \\
  \midrule
  \multirow{6}{*}{(a)} & DeepWalk & $0.866$ & $0.889$ & $0.86$ \\
  & COSINE-DW & $\textbf{0.896}$ & $\textbf{0.932}$ & $\textbf{0.881}$ \\
  \cmidrule{2-5}
  & node2vec & $0.876$ & $0.889$ & $0.852$ \\
  & COSINE-N2V & $\textbf{0.900}$ & $\textbf{0.936}$ & $\textbf{0.888}$\\
  \cmidrule{2-5}
  & LINE$_{2nd}$ & $0.620$ & $0.817$ & $0.626$ \\
  & COSINE-LINE$_{2nd}$ & $\textbf{0.775}$ & $\textbf{0.865}$ & $\textbf{0.746}$ \\
  \midrule
  \multirow{6}{*}{(b)} & DeepWalk & $0.874$ & $0.893$ & $0.867$ \\
  & COSINE-DW & $\textbf{0.898}$ & $\textbf{0.930}$ & $\textbf{0.887}$ \\
  \cmidrule{2-5}
  & node2vec & $0.880$ & $0.894$ & $0.858$ \\
  & COSINE-N2V & $\textbf{0.899}$ & $\textbf{0.935}$ & $\textbf{0.894}$ \\
  \cmidrule{2-5}
  & LINE$_{2nd}$ & $0.629$ & $0.823$ & $0.635$ \\
  & COSINE-LINE$_{2nd}$ & $\textbf{0.731}$ & $\textbf{0.836}$ & $\textbf{0.716}$ \\
  \midrule
  \multirow{6}{*}{(c)} & DeepWalk & $0.926$ & $0.927$ & $0.943$ \\
  & COSINE-DW & $\textbf{0.941}$ & $\textbf{0.968}$ & $\textbf{0.951}$ \\
  \cmidrule{2-5}
  & node2vec & $0.926$ & $0.928$ & $0.945$ \\
  & COSINE-N2V & $\textbf{0.942}$ & $\textbf{0.971}$ & $\textbf{0.953}$ \\
  \cmidrule{2-5}
  & LINE$_{2nd}$ & $0.921$ & $0.934$ & $0.943$ \\
  & COSINE-LINE$_{2nd}$ & $\textbf{0.962}$ & $\textbf{0.963}$ & $\textbf{0.956}$ \\
  \midrule

  \end{tabular}

  \end{center}
\end{table}

\begin{table}[!htb]
  \caption{Mean Reciprocal Rank (MRR) scores for link prediction. Comparison between compressed and uncompressed models using binary operators: (a) L1-norm, (b) L2-norm, and (c) Dot product.} 
  \label{table:result-lp-mrr}
  \begin{center}
  \begin{tabular}{c|c|c|c|c}
  \multirow{2}{*}{\textbf{Op}} & \multirow{2}{*}{\textbf{Algorithm}} & \multicolumn{3}{c}{\textbf{Dataset}} \\
  & & Youtube & Flickr & Yelp \\
  \midrule
  \multirow{6}{*}{(a)} & DeepWalk & $0.828$ & $0.855$ & $0.743$ \\
  & COSINE-DW & $\textbf{0.851}$ & $\textbf{0.898}$ & $\textbf{0.766}$ \\
  \cmidrule{2-5}
  & node2vec & $0.838$ & $0.854$ & $0.737$ \\
  & COSINE-N2V & $\textbf{0.853}$ & $\textbf{0.902}$ & $\textbf{0.775}$ \\
  \cmidrule{2-5}
  & LINE$_{2nd}$ & $0.628$ & $0.801$ & $0.572$ \\
  & COSINE-LINE$_{2nd}$ & $\textbf{0.746}$ & $\textbf{0.830}$ & $\textbf{0.650}$ \\
  \midrule
  \multirow{6}{*}{(b)} & DeepWalk & $0.839$ & $0.859$ & $0.754$ \\
  & COSINE-DW & $\textbf{0.851}$ & $\textbf{0.893}$ & $\textbf{0.771}$ \\
  \cmidrule{2-5}
  & node2vec & $0.845$ & $0.861$ & $0.749$ \\
  & COSINE-N2V & $\textbf{0.852}$ & $\textbf{0.898}$ & $\textbf{0.782}$\\
  \cmidrule{2-5}
  & LINE$_{2nd}$ & $0.639$ & $0.809$ & $0.586$ \\
  & COSINE-LINE$_{2nd}$ & $\textbf{0.700}$ & $\textbf{0.819}$ & $\textbf{0.606}$ \\
  \midrule
  \multirow{6}{*}{(c)} & DeepWalk & $0.874$ & $0.874$ & $0.85$ \\
  & COSINE-DW & $\textbf{0.905}$ & $\textbf{0.946}$ & $\textbf{0.876}$ \\
  \cmidrule{2-5}
  & node2vec & $0.874$ & $0.876$ & $0.857$ \\
  & COSINE-N2V & $\textbf{0.906}$ & $\textbf{0.950}$ & $\textbf{0.882}$\\
  \cmidrule{2-5}
  & LINE$_{2nd}$ & $0.875$ & $0.905$ & $0.859$ \\
  & COSINE-LINE$_{2nd}$ & $\textbf{0.939}$ & $\textbf{0.935}$ & $\textbf{0.892}$ \\
  \midrule

  \end{tabular}

  \end{center}
\end{table}

\subsection{Multi-Label Classification}

\begin{table*}[!htb]
  \caption{Multi-label classification results in Youtube.}
  \label{table:result-clf-youtube}
  \begin{center}
  \begin{tabular}{c|c|c|c|c|c|c|c|c|c|c|c}
  \toprule
  & \%Training ratio & $1\%$ & $2\%$ & $3\%$ & $4\%$ & $5\%$ & $6\%$ & $7\%$ & $8\%$ & $9\%$ & $10\%$ \\
  \midrule
  \multirow{6}{*}{Micro-F1(\%)} & DeepWalk & $31.1\%$ & $33.3\%$ & $34.8\%$ & $35.9\%$ & $36.2\%$  & $36.7\%$ & $36.8\%$ & $37.0\%$ & $37.3\%$ & $37.4\%$ \\
  & COSINE-DW & $\textbf{36.5\%}$ & $\textbf{39.5\%}$ & $\textbf{41.1\%}$ & $\textbf{42.0\%}$ & $\textbf{42.3\%}$ & $\textbf{42.8\%}$ & $\textbf{43.3\%}$ & $\textbf{43.6\%}$ & $\textbf{43.9\%}$ & $\textbf{44.0\%}$ \\
  \cmidrule{2-12}
  & node2vec($p=2,q=2$) & $31.3\%$ & $33.2\%$ & $35.0\%$ & $36.5\%$ & $36.8\%$ & $37.4\%$ & $37.4\%$ & $37.7\%$ & $37.9\%$ & $38.0\%$ \\
  & COSINE-N2V($p=0.25,q=0.5$) & $\textbf{36.6\%}$ & $\textbf{39.4\%}$ & $\textbf{40.8\%}$ & $\textbf{41.8\%}$ & $\textbf{42.2\%}$ & $\textbf{42.6\%}$ & $\textbf{43.1\%}$ & $\textbf{43.6\%}$ & $\textbf{44.0\%}$ & $\textbf{44.1\%}$ \\
  \cmidrule{2-12}
    & LINE$_{2nd}$ & $30.9\%$ & $32.7\%$ & $34.0\%$ & $34.7\%$ & $35.1\%$ & $35.7\%$ & $35.9\%$ & $36.1\%$ & $36.3\%$ & $36.2\%$ \\
  & COSINE-LINE$_{2nd}$ & $\textbf{36.3\%}$ & $\textbf{39.8\%}$ & $\textbf{41.6\%}$ & $\textbf{42.4\%}$ & $\textbf{42.7\%}$ & $\textbf{43.2\%}$ & $\textbf{43.6\%}$ & $\textbf{43.9\%}$ & $\textbf{44.4\%}$ & $\textbf{44.4\%}$ \\
  \midrule
    \multirow{6}{*}{Macro-F1(\%)} & DeepWalk & $14.0\%$ & $16.2\%$ & $17.8\%$ & $20.4\%$ & $21.0\%$ & $22.0\%$ & $22.5\%$ & $23.0\%$ & $23.5\%$ & $23.8\%$ \\
  & COSINE-DW & $\textbf{21.2\%}$ & $\textbf{24.6\%}$ & $\textbf{27.5\%}$ & $\textbf{29.4\%}$ & $\textbf{30.0\%}$ & $\textbf{30.7\%}$ & $\textbf{31.7\%}$ & $\textbf{32.1\%}$ & $\textbf{32.8\%}$ & $\textbf{32.9\%}$ \\
  \cmidrule{2-12}
  & node2vec($p=2,q=2$) & $14.3\%$ & $16.3\%$ & $18.5\%$ & $21.0\%$ & $21.6\%$ & $22.7\%$ & $22.9\%$ & $23.5\%$ & $24.0\%$ & $24.2\%$ \\
  & COSINE-N2V($p=0.25,q=0.5$) & $\textbf{21.2\%}$ & $\textbf{24.2\%}$ & $\textbf{27.2\%}$ & $\textbf{29.2\%}$ & $\textbf{29.9\%}$ & $\textbf{30.6\%}$ & $\textbf{31.7\%}$ & $\textbf{32.4\%}$ & $\textbf{33.2\%}$ & $\textbf{33.2\%}$ \\
  \cmidrule{2-12}
    & LINE$_{2nd}$ & $14.1\%$ & $16.3\%$ & $18.3\%$ & $20.4\%$ & $21.1\%$ & $22.1\%$ & $22.5\%$ & $22.9\%$ & $23.4\%$ & $23.4\%$ \\
  & COSINE-LINE$_{2nd}$ & $\textbf{21.4\%}$ & $\textbf{26.0\%}$ & $\textbf{29.3\%}$ & $\textbf{31.3\%}$ & $\textbf{31.7\%}$ & $\textbf{32.7\%}$ & $\textbf{33.7\%}$ & $\textbf{34.1\%}$ & $\textbf{34.9\%}$ & $\textbf{35.0\%}$ \\
  \bottomrule
  \end{tabular}
  \end{center}
\end{table*}

\begin{table*}[!htb]
  \caption{Multi-label classification results in Flickr.}
  \label{table:result-clf-flickr}
  \begin{center}
  \begin{tabular}{c|c|c|c|c|c|c|c|c|c|c|c}
  \toprule
  & \%Training ratio & $1\%$ & $2\%$ & $3\%$ & $4\%$ & $5\%$ & $6\%$ & $7\%$ & $8\%$ & $9\%$ & $10\%$ \\
  \midrule
  \multirow{6}{*}{Micro-F1(\%)} & DeepWalk & $39.7\%$ & $40.2\%$ & $40.4\%$ & $40.6\%$ & $40.8\%$ & $40.8\%$ & $40.9\%$ & $40.9\%$ & $41.0\%$ & $41.0\%$ \\
  & COSINE-DW & $\textbf{40.4\%}$ & $\textbf{41.0\%}$ & $\textbf{41.4\%}$ & $\textbf{41.6\%}$ & $\textbf{41.9\%}$ & $\textbf{42.0\%}$ & $\textbf{42.1\%}$ & $\textbf{42.2\%}$ & $\textbf{42.2\%}$ & $\textbf{42.3\%}$ \\
  \cmidrule{2-12}
  & node2vec($p=2,q=0.5$) & $39.8\%$ & $40.2\%$ & $40.5\%$ & $40.7\%$ & $40.8\%$ & $40.9\%$ & $40.9\%$ & $41.0\%$ & $41.0\%$ & $41.0\%$ \\ 
  & COSINE-N2V($p=1,q=1$) & $\textbf{40.4\%}$ & $\textbf{41.0\%}$ & $\textbf{41.4\%}$ & $\textbf{41.6\%}$ & $\textbf{41.9\%}$ & $\textbf{42.0\%}$ & $\textbf{42.1\%}$ & $\textbf{42.2\%}$ & $\textbf{42.2\%}$ & $\textbf{42.3\%}$ \\
  \cmidrule{2-12}
    & LINE$_{2nd}$ & $\textbf{41.0\%}$ & $41.3\%$ & $41.5\%$ & $41.7\%$ & $41.8\%$ & $41.8\%$ & $41.8\%$ & $41.9\%$ & $41.9\%$ & $41.9\%$ \\
  & COSINE-LINE$_{2nd}$ & $40.8\%$ & $\textbf{41.4\%}$ & $\textbf{41.8\%}$ & $\textbf{42.1\%}$ & $\textbf{42.4\%}$ & $\textbf{42.6\%}$ & $\textbf{42.7\%}$ & $\textbf{42.8\%}$ & $\textbf{42.9\%}$ & $\textbf{42.9\%}$ \\
  \midrule
    \multirow{6}{*}{Macro-F1(\%)} & DeepWalk & $26.8\%$ & $28.2\%$ & $29.3\%$ & $29.9\%$ & $30.3\%$ & $30.4\%$ & $30.6\%$ & $30.8\%$ & $31.0\%$ & $31.0\%$ \\
  & COSINE-DW & $\textbf{29.7\%}$ & $\textbf{31.4\%}$ & $\textbf{32.9\%}$ & $\textbf{33.6\%}$ & $\textbf{34.1\%}$ & $\textbf{34.2\%}$ & $\textbf{34.4\%}$ & $\textbf{34.6\%}$ & $\textbf{34.7\%}$ & $\textbf{34.9\%}$ \\
  \cmidrule{2-12}
  & node2vec($p=2,q=0.5$) & $27.1\%$ & $28.3\%$ & $29.4\%$ & $30.1\%$ & $30.5\%$ & $30.6\%$ & $30.8\%$ & $31.0\%$ & $31.1\%$ & $31.2\%$ \\
  & COSINE-N2V($p=1,q=1$) & $\textbf{29.7\%}$ & $\textbf{31.4\%}$ & $\textbf{32.9\%}$ & $\textbf{33.6\%}$ & $\textbf{34.1\%}$ & $\textbf{34.2\%}$ & $\textbf{34.4\%}$ & $\textbf{34.6\%}$ & $\textbf{34.7\%}$ & $\textbf{34.9\%}$ \\
  \cmidrule{2-12}
    & LINE$_{2nd}$ & $30.1\%$ & $31.3\%$ & $32.1\%$ & $32.8\%$ & $33.0\%$ & $33.1\%$ & $33.2\%$ & $33.3\%$ & $33.3\%$ & $33.4\%$ \\
  & COSINE-LINE$_{2nd}$ & $\textbf{32.0\%}$ & $\textbf{33.6\%}$ & $\textbf{34.8\%}$ & $\textbf{35.5\%}$ & $\textbf{35.9\%}$ & $\textbf{36.1\%}$ & $\textbf{36.2\%}$ & $\textbf{36.4\%}$ & $\textbf{36.5\%}$ & $\textbf{36.6\%}$ \\
  \bottomrule
  \end{tabular}
  \end{center}
\end{table*}

\begin{table*}[!htb]
  \caption{Multi-label classification results in Yelp.}
  \label{table:result-clf-yelp}
  \begin{center}
  \begin{tabular}{c|c|c|c|c|c|c|c|c|c|c|c}
  \toprule
  & \%Training ratio & $1\%$ & $2\%$ & $3\%$ & $4\%$ & $5\%$ & $6\%$ & $7\%$ & $8\%$ & $9\%$ & $10\%$ \\
  \midrule
  \multirow{6}{*}{Micro-F1(\%)} & DeepWalk & $63.2\%$ & $63.2\%$ & $63.3\%$ & $63.3\%$ & $63.3\%$ & $63.3\%$ & $63.3\%$ & $63.3\%$ & $63.3\%$ & $63.3\%$ \\
  & COSINE-DW & $\textbf{63.4\%}$ & $\textbf{63.6\%}$ & $\textbf{63.7\%}$ & $\textbf{63.8\%}$ & $\textbf{63.9\%}$ & $\textbf{63.9\%}$ & $\textbf{64.0\%}$ & $\textbf{64.0\%}$ & $\textbf{64.0\%}$ & $\textbf{64.0\%}$ \\  \cmidrule{2-12}
  & node2vec($p=0.5,q=2$) & $63.3\%$ & $63.4\%$ & $63.4\%$ & $63.4\%$ & $63.4\%$ & $63.4\%$ & $63.4\%$ & $63.4\%$ & $63.4\%$ & $63.4\%$ \\
  & COSINE-N2V($p=0.5,q=2$) & $\textbf{63.4\%}$ & $\textbf{63.7\%}$ & $\textbf{63.8\%}$ & $\textbf{63.8\%}$ & $\textbf{63.9\%}$ & $\textbf{63.9\%}$ & $\textbf{63.9\%}$ & $\textbf{64.0\%}$ & $\textbf{64.0\%}$ & $\textbf{64.0\%}$ \\
  \cmidrule{2-12}
    & LINE$_{2nd}$ & $63.2\%$ & $63.2\%$ & $63.3\%$ & $63.3\%$ & $63.3\%$ & $63.3\%$ & $63.3\%$ & $63.3\%$ & $63.2\%$ & $63.3\%$ \\
  & COSINE-LINE$_{2nd}$ & $\textbf{63.4\%}$ & $\textbf{63.6\%}$ & $\textbf{63.6\%}$ & $\textbf{63.7\%}$ & $\textbf{63.7\%}$ & $\textbf{63.8\%}$ & $\textbf{63.8\%}$ & $\textbf{63.8\%}$ & $\textbf{63.8\%}$ & $\textbf{63.8\%}$ \\
  \midrule
    \multirow{6}{*}{Macro-F1(\%)} & DeepWalk & $34.6\%$ & $34.6\%$ & $34.7\%$ & $34.8\%$ & $34.8\%$ & $34.8\%$ & $34.8\%$ & $34.8\%$ & $34.8\%$ & $34.8\%$ \\
  & COSINE-DW & $\textbf{36.0\%}$ & $\textbf{36.2\%}$ & $\textbf{36.3\%}$ & $\textbf{36.4\%}$ & $\textbf{36.4\%}$ & $\textbf{36.5\%}$ & $\textbf{36.5\%}$ & $\textbf{36.5\%}$ & $\textbf{36.4\%}$ & $\textbf{36.4\%}$ \\
  \cmidrule{2-12}
  & node2vec($p=0.5,q=2$) & $35.0\%$ & $34.9\%$ & $35.0\%$ & $35.1\%$ & $35.1\%$ & $35.1\%$ & $35.1\%$ & $35.1\%$ & $35.1\%$ & $35.1\%$ \\
  & COSINE-N2V($p=0.5,q=2$) & $\textbf{36.1\%}$ & $\textbf{36.2\%}$ & $\textbf{36.3\%}$ & $\textbf{36.4\%}$ & $\textbf{36.5\%}$ & $\textbf{36.5\%}$ & $\textbf{36.5\%}$ & $\textbf{36.5\%}$ & $\textbf{36.4\%}$ & $\textbf{36.5\%}$ \\
  \cmidrule{2-12}
    & LINE$_{2nd}$ & $35.1\%$ & $35.1\%$ & $35.2\%$ & $35.2\%$ & $35.2\%$ & $35.3\%$ & $35.3\%$ & $35.2\%$ & $35.3\%$ & $35.3\%$ \\
  & COSINE-LINE$_{2nd}$ & $\textbf{36.0\%}$ & $\textbf{35.9\%}$ & $\textbf{36.1\%}$ & $\textbf{36.2\%}$ & $\textbf{36.3\%}$ & $\textbf{36.3\%}$ & $\textbf{36.3\%}$ & $\textbf{36.3\%}$ & $\textbf{36.2\%}$ & $\textbf{36.2\%}$ \\
  \bottomrule
  \end{tabular}
  \end{center}
\end{table*}

For multi-label classification task, we randomly select a portion of nodes as the training set and leave the rest as the test set. We treat network embeddings as nodes' features and feed them into a one-vs-rest SVM classifier implemented by \textit{LibLinear}~\cite{fan2008liblinear} as previous works did~\cite{yang2017fast}. Consider classification in large networks, where there is a little portion of labeled nodes. We vary the training ratio from $1\%$ to $10\%$ to see the performance under sparse situation. To avoid overfitting, we train the classifier with L2-regularization.

We report the results under the best number of samples and compare the capacity in Table \ref{table:result-clf-youtube}, Table \ref{table:result-clf-flickr} and Table \ref{table:result-clf-yelp}. Numbers in bold represent the higher performance between compressed and uncompressed models. From these tables, we have the following observations:
\begin{enumerate}[(1)]
\item The proposed COSINE framework consistently and significantly improves all baseline methods on node classification. In Youtube, COSINE gives us at least $13\%$ gain over all baselines in Micro-F1 and gives us at least $24\%$ gain over all baselines in Macro-F1. In case of Flickr network, COSINE gives us at least $2\%$ gain over all baselines in Micro-F1 and gives us at least $6\%$ gain over all baselines in Macro-F1. As we can see in Yelp network, the classification scores don't change a lot with the growing training ratio, which can be explained by the weak relation between network structure information and nodes' labels. The link prediction result has proven COSINE can guarantee the high-quality of network embeddings in Yelp dataset. Although there is little label information in network structure, COSINE still helps baselines extract it better.
\item LINE$_{2nd}$ just consider the second order proximity in networks. As the previous work showed~\cite{dalmia2018towards}, LINE$_2nd$ has bad performance when the network is sparse like Youtube compared to DeepWalk and node2vec. COSINE encode high order proximity before training, which helps LINE$_2nd$ achieve the comparative performance in sparse networks.
\item For node2vec with COSINE, the best return parameter $q$ is not more than 1 in all networks, which means the local structure is less useful for the model training. Nodes in the same local structure share part of parameters while training. So, there is no need to revisit neighbor nodes, which is consistent with our framework design.
\end{enumerate}
To summarise, COSINE framework effectively encodes high order proximity before training, which is crucial for the parameter sharing. And the parameter sharing improve the capacity of baselines under the limited memory. Besides, COSINE is flexible to various social networks, whether they are sparse or dense. Moreover, it is a general framework, which works well with all baseline methods.

\subsection{Scalability}

\begin{figure*}[!htb]
\centering
\subfigure[]{                  
\begin{minipage}{0.24\linewidth}
\centering
\includegraphics[scale=0.23, trim={0 7cm 0 5cm}]{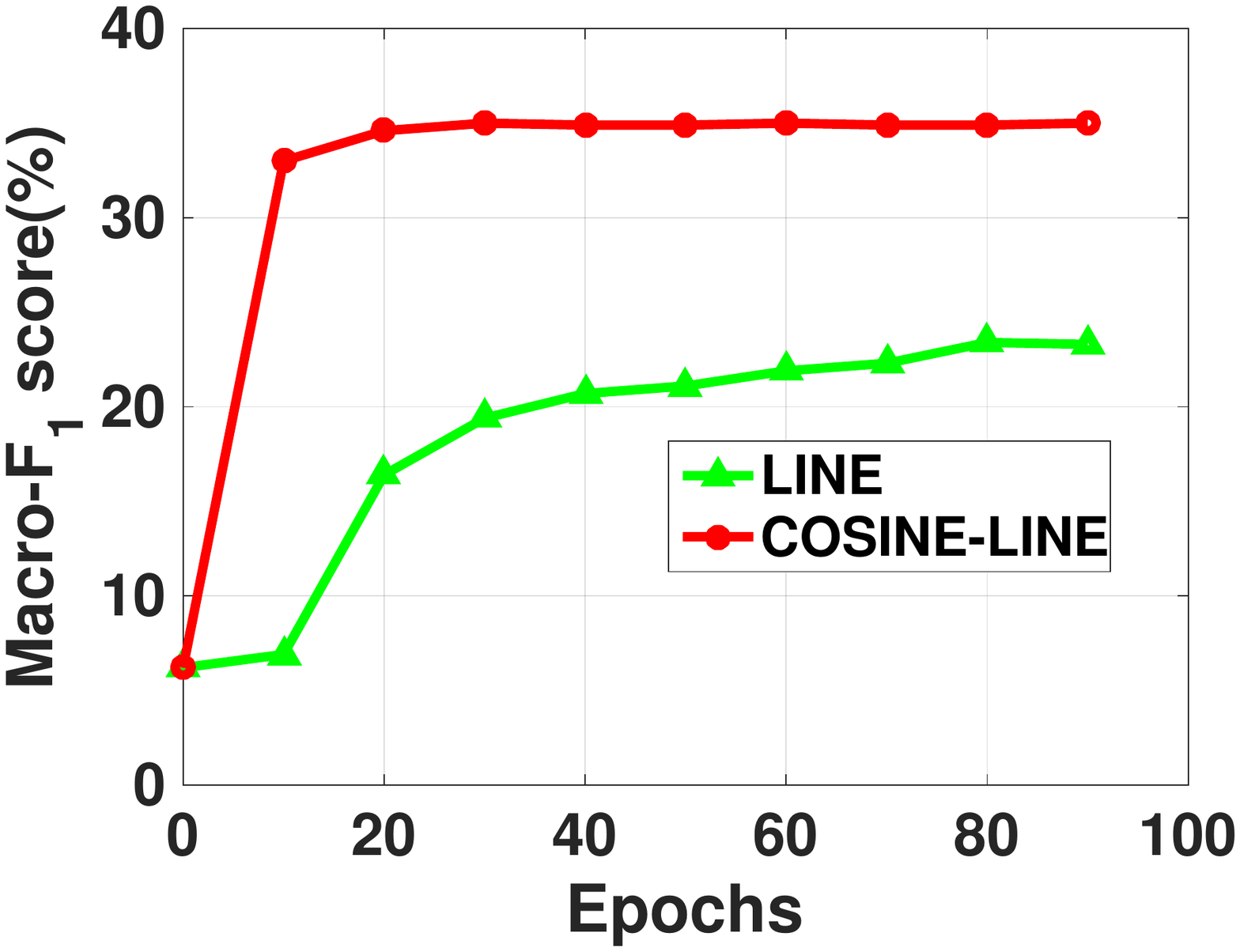}
\label{fig:3:a}         
\end{minipage}}
\subfigure[]{                    
\begin{minipage}{0.24\linewidth}
\centering
\includegraphics[scale=0.23, trim={0 7cm 0 5cm}]{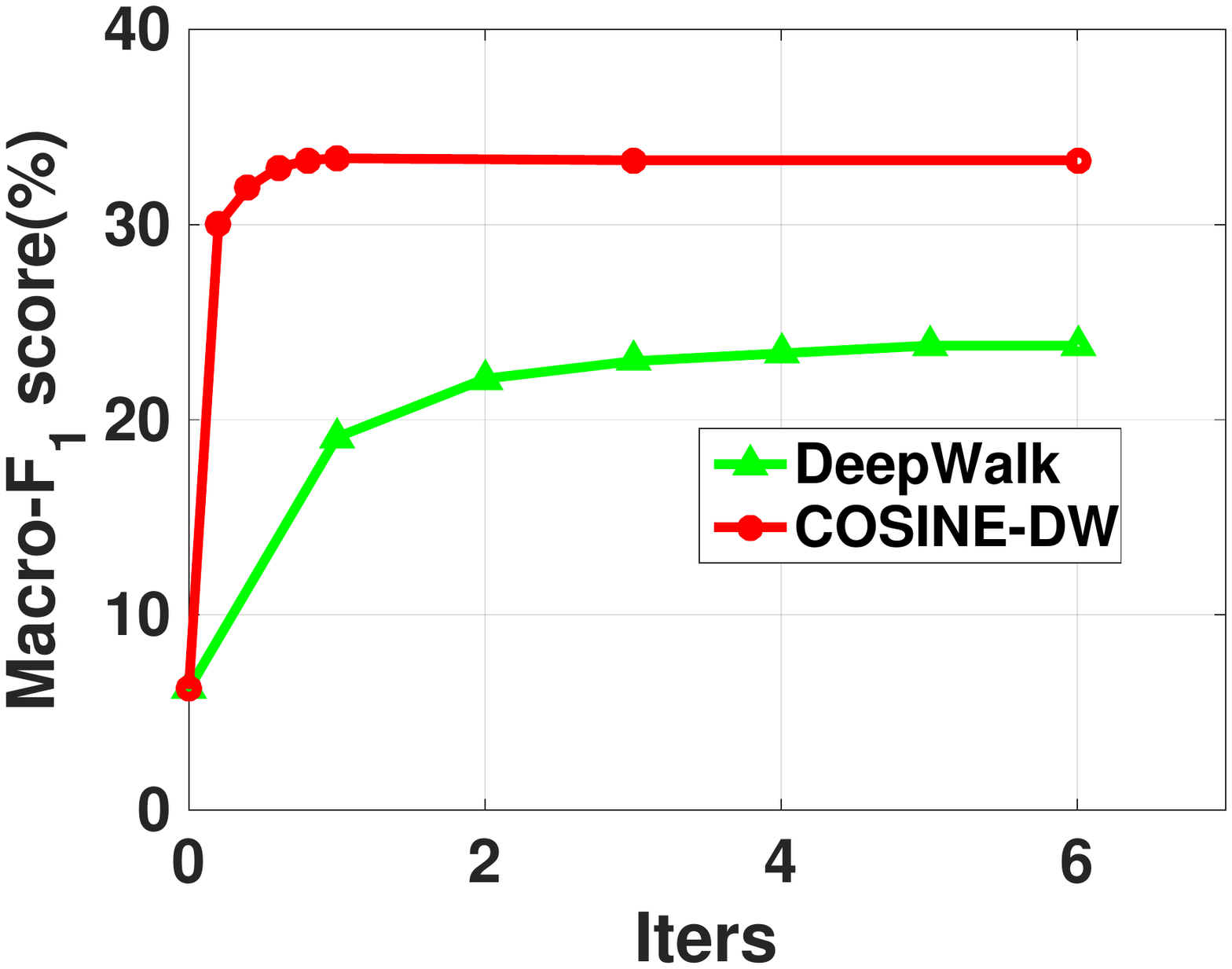}
\label{fig:3:b}             
\end{minipage}}
\subfigure[]{                    
\begin{minipage}{0.24\linewidth}
\centering
\includegraphics[scale=0.23, trim={0 7cm 0 5cm}]{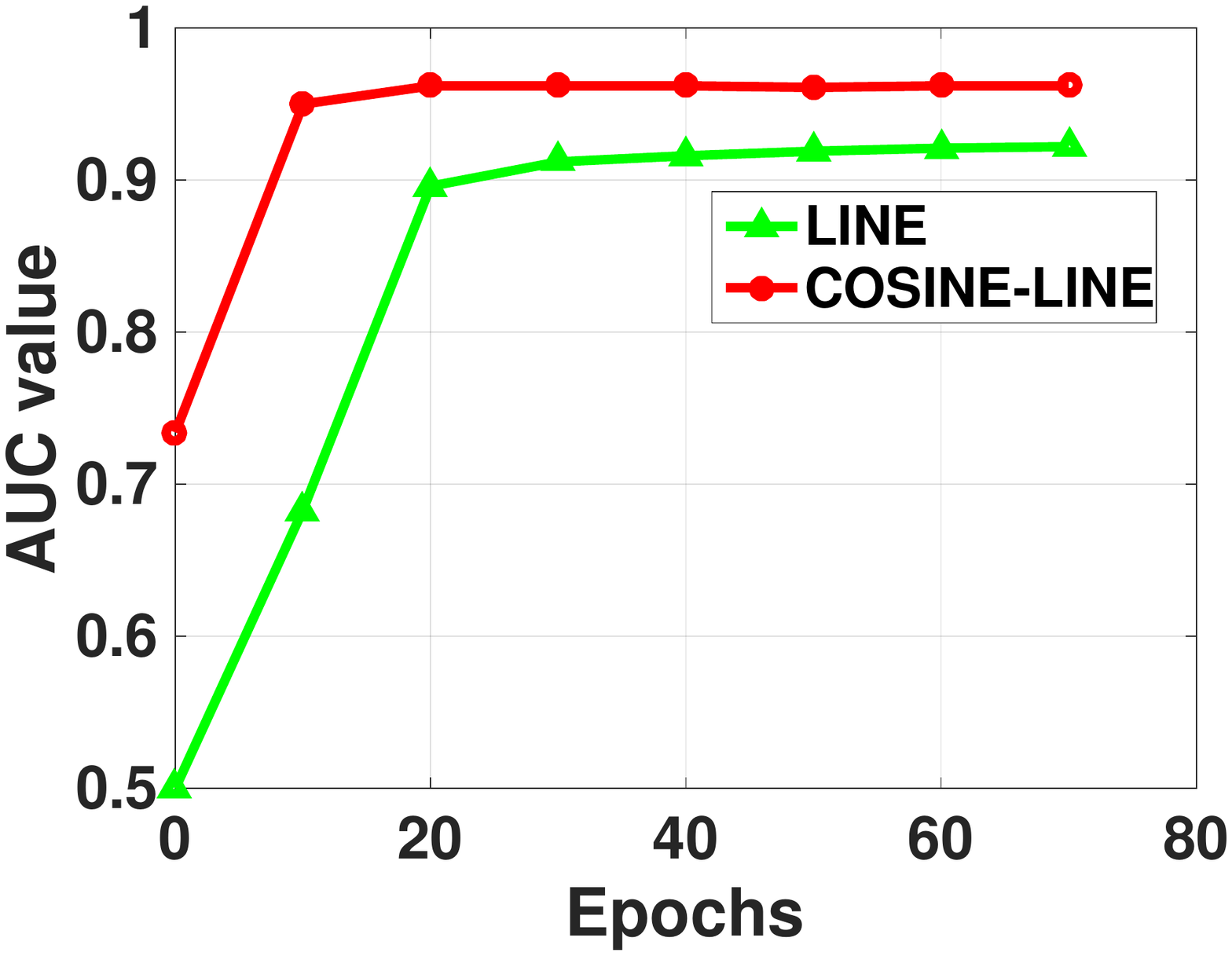}
\label{fig:3:c}             
\end{minipage}}
\subfigure[]{                    
\begin{minipage}{0.24\linewidth}
\centering
\includegraphics[scale=0.23, trim={0 7cm 0 5cm}]{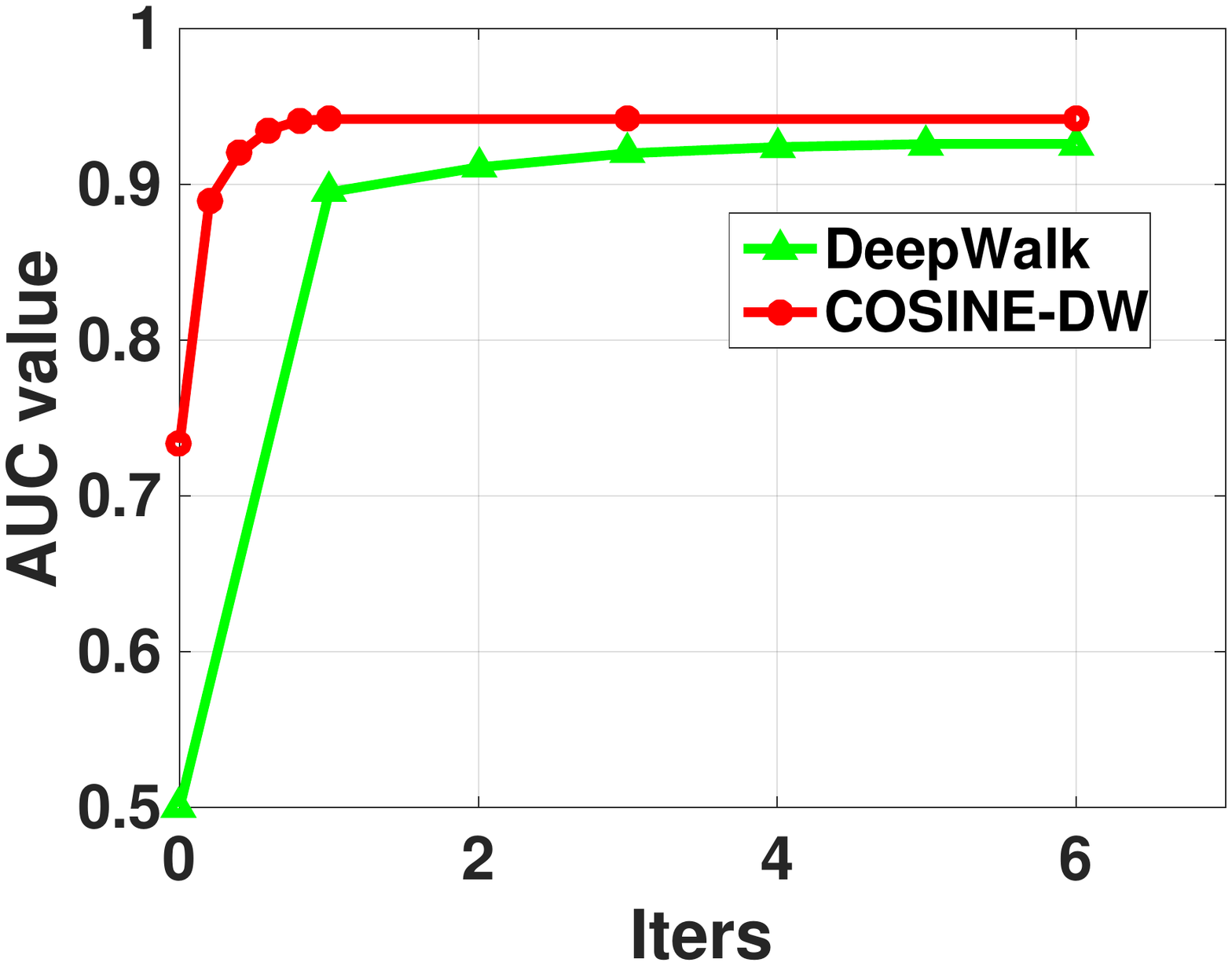}
\label{fig:3:d}             
\end{minipage}}
\caption{(a) (b) show the classification performance w.r.t. the number of samples and (c) (d) show the link prediction performance w.r.t. the number of samples. The definitions of epoch and iteration are in Section 4.3, and the dataset is Youtube.}                  
\label{fig:3}                                                     
\end{figure*}

\begin{figure}[!htb]
\centering
\includegraphics[scale=0.4, trim={0 7cm 0 7cm}]{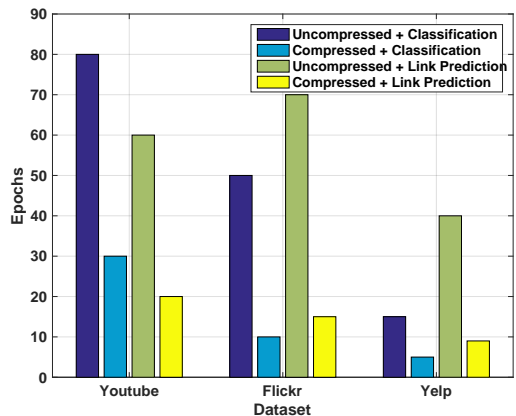}
\caption{The best number of training epochs for LINE w.r.t. different tasks and datasets.}
\label{fig:4}                                                     
\end{figure}

\begin{figure}[!htb]
\centering
\includegraphics[scale=0.4, trim={0 7cm 0 7cm}]{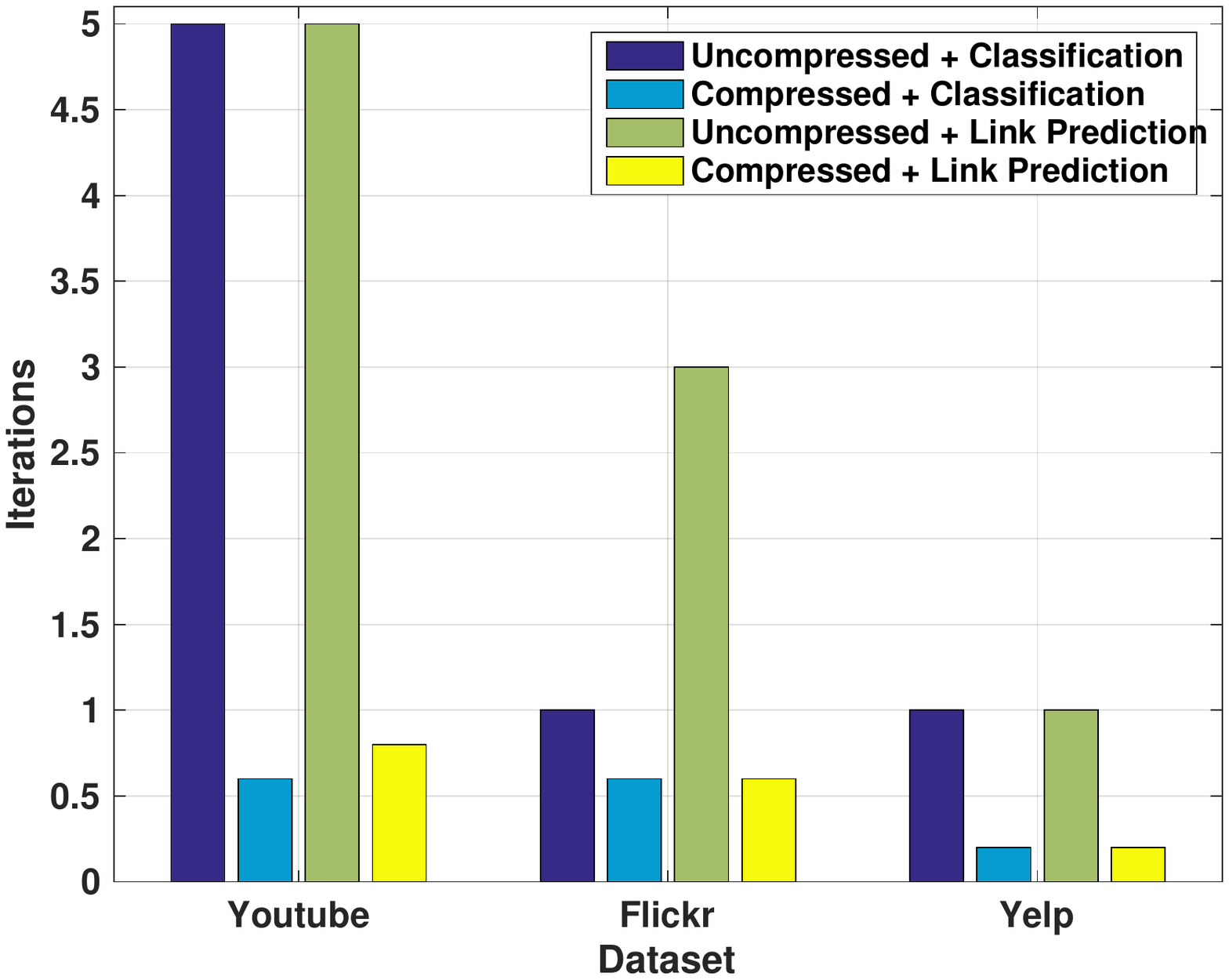}
\caption{The best number of training iterations for DeepWalk w.r.t. different tasks and datasets.}
\label{fig:5}                            
\end{figure}

We now explore the scalability of our COSINE framework on three large-scale networks. As mentioned earlier, we should find the best value of sample number which makes the model train to convergence. The scalable model needs to take fewer samples to achieve better performance. In Fig. \ref{fig:3}, we report the classification performance w.r.t. the training samples with $10\%$ training ratio and the link prediction performance w.r.t. the training samples with dot product score function on Youtube network.

Fig. \ref{fig:3:a} and Fig. \ref{fig:3:b} show that the classification performances of compressed and uncompressed models are the same without training data, which means graph partitioning results are useless at the beginning. Although the start Macro-F$_1$ scores are the same, the score of compressed models grows faster with graph partitioning. Besides, compressed models achieve convergence before uncompressed models. Moreover, compressed models just use little training data to outperform uncompressed models with convergence, i.e., COSINE-LINE with 10 epochs gives us 40\% gain over LINE with 90 epochs.

Fig. \ref{fig:3:c} and Fig. \ref{fig:3:d} show that the link prediction performances of compressed and uncompressed models are different without training data, which means graph partitioning results are useful at the beginning. As the result of classification shows, compressed models also achieve convergence before uncompressed models while the growing speeds are nearly the same.

For three datasets, we summary the best values of samples for LINE and DeepWalk in Fig. \ref{fig:4} and \ref{fig:5}. We treat the best iteration number for DeepWalk as the best one for node2vec in that they are both random-walk based methods and it is expensive for node2vec to find best iteration number when employing grid searching for hyperparameters $p$ and $q$. From these two figures, we underline the following observations:
\begin{enumerate}[(1)]
\item Compressed models consistently and significantly reduce the training samples compared to the original models. It states the importance of incorporating graph partitioning results before training. Graph partitioning can encode high order proximity of the network, which is hard for baseline methods to learn.
\item For models with COSINE framework, the best numbers of samples for two evaluation tasks are very close in a specific dataset while the best numbers are sometimes quite different for uncompressed models. It indicates that COSINE improves the stability of models among different tasks.
\end{enumerate}

In a word, graph partitioning and parameters sharing help baselines to learn faster from data and reduce the need for training samples. We will discuss the time cost of graph partitioning and parameters sharing later in detail to show the time efficiency of COSINE.

\subsection{Time Efficiency}

\begin{figure*}[!htb]
\centering
\subfigure[LINE's time w.r.t. datasets]{                  
\begin{minipage}{0.3\linewidth}
\centering
\includegraphics[scale=0.32, trim={4cm 7cm 0 3cm}]{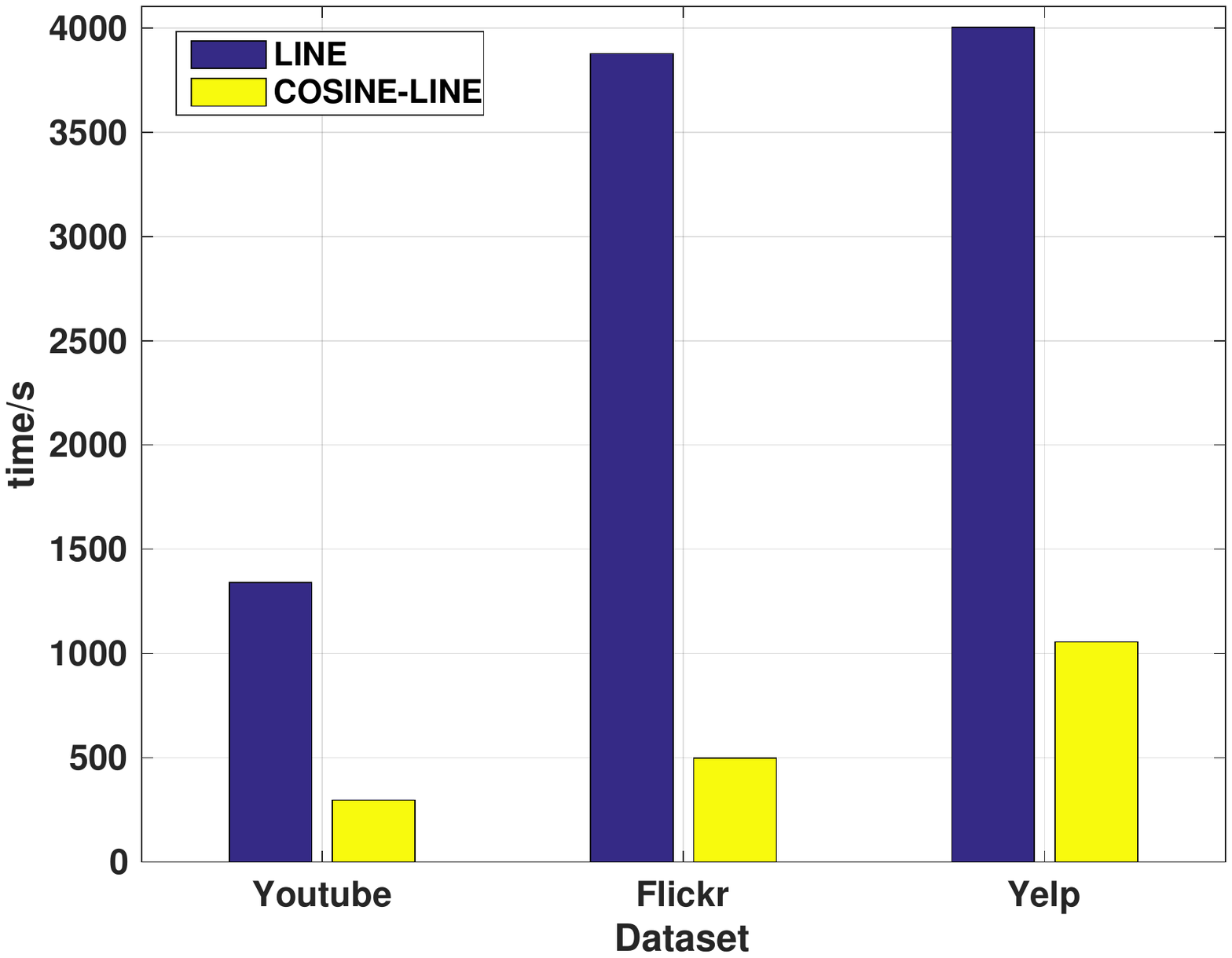}
\label{fig:8:a}         
\end{minipage}}
\subfigure[DeepWalk's time w.r.t. datasets]{                    
\begin{minipage}{0.3\linewidth}
\centering
\includegraphics[scale=0.32, trim={3cm 7cm 0 3cm}]{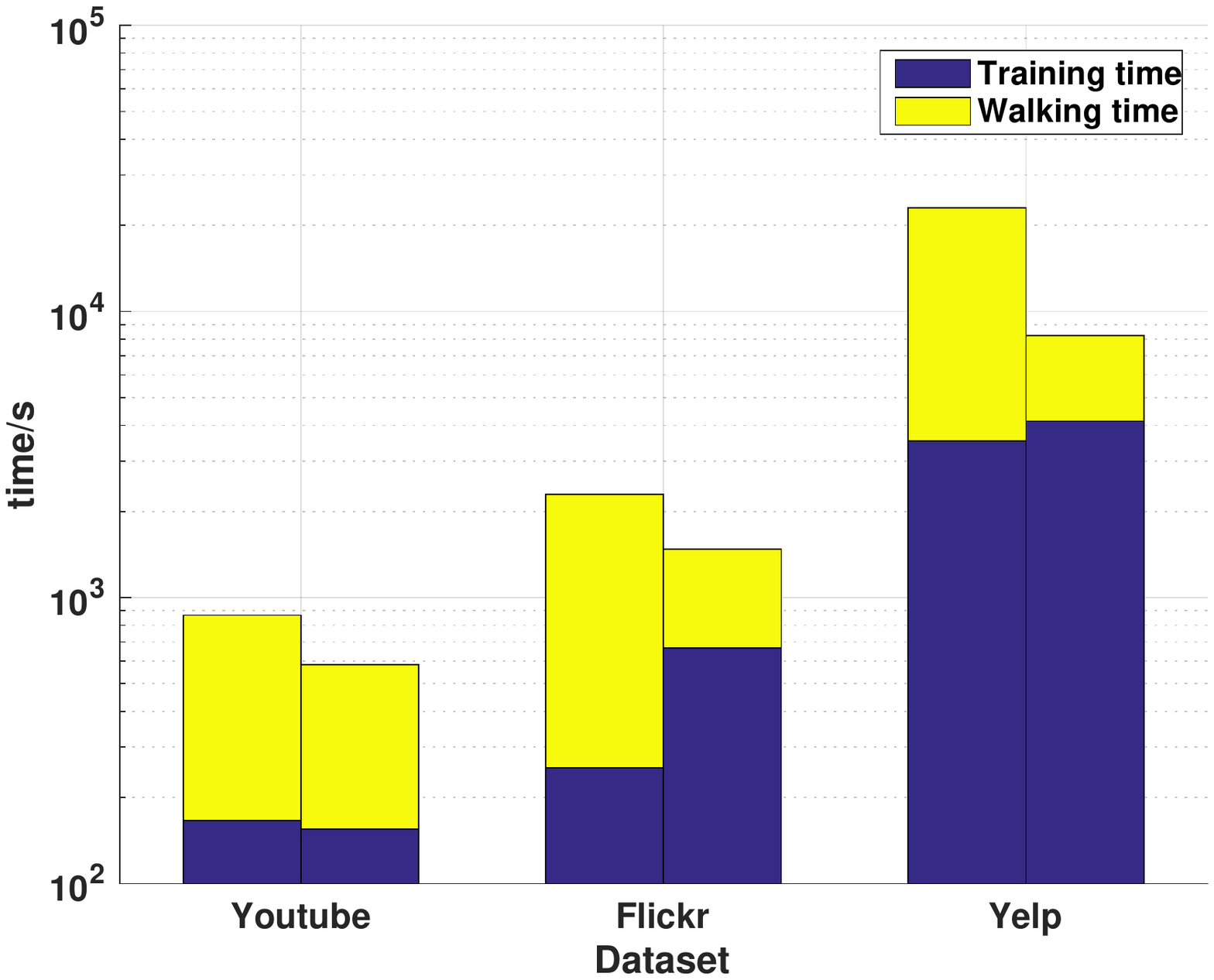}
\label{fig:8:b}             
\end{minipage}}
\subfigure[Time w.r.t. datasets]{                    
\begin{minipage}{0.3\linewidth}
\centering
\includegraphics[scale=0.32, trim={2cm 7cm 0 3cm}]{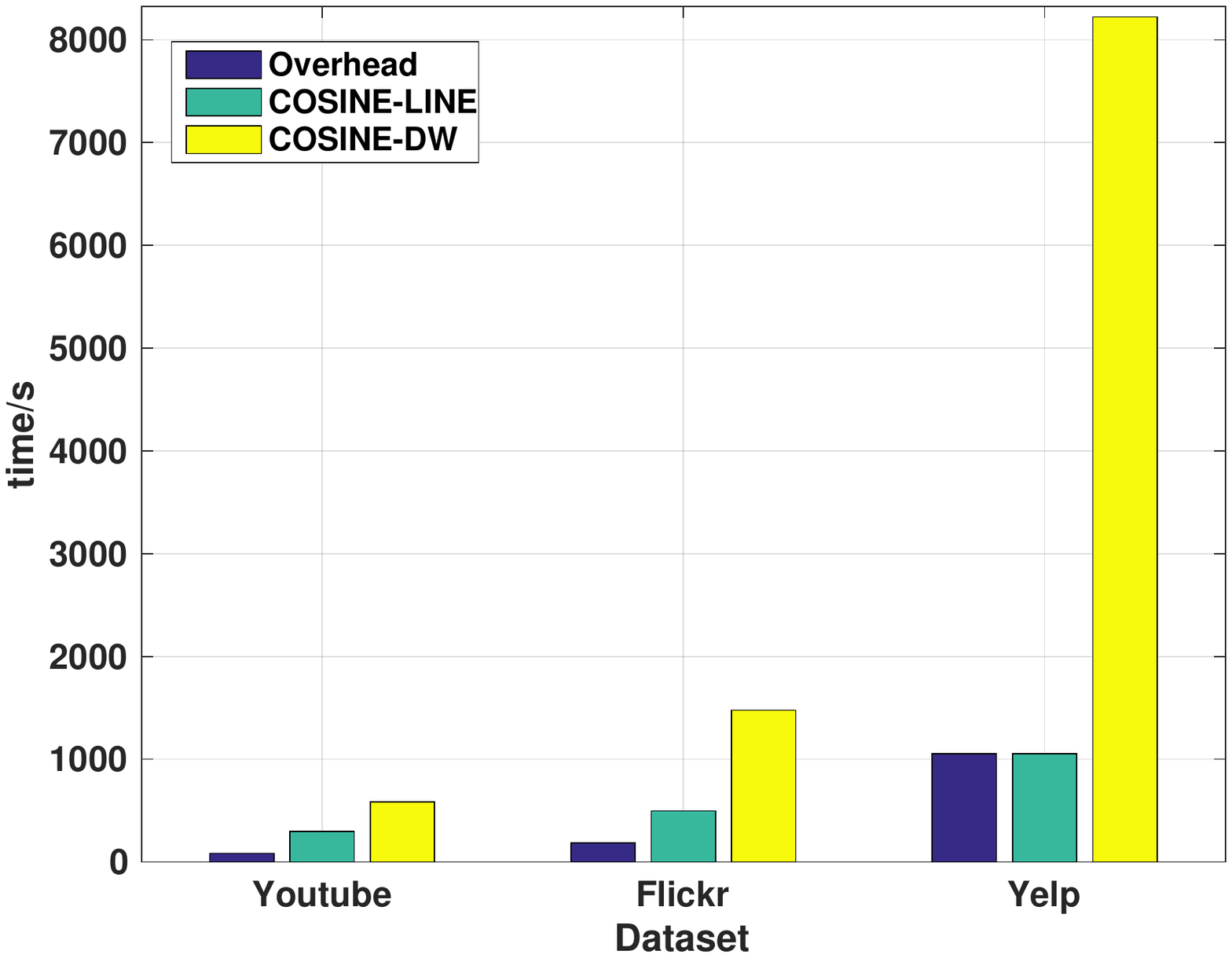}
\label{fig:8:c}             
\end{minipage}}
\caption{Time performance on three large-scale networks. Note that the left part is DeepWalk and the right part is COSINE-DW in (b).} 
\label{fig:8}
\end{figure*}

In this subsection, we explore the time efficiency of our COSINE framework on three large-scale networks.
We conduct network embedding on a modern machine with 12 cores.
Fig. \ref{fig:8:a} shows the running time to convergence of compressed and uncompressed models about LINE.
We observe that COSINE significantly and consistently reduces the running time on three datasets by at least 20\%.
Fig. \ref{fig:8:b} show the running time convergence of compressed and uncompressed models for DeepWalk.
There are training time and walking time for DeepWalk. As COSINE-DW need less walks for training to convergence, the walking times of COSINE-DW are also less than those of DeepWalk.
COSINE also reduces the running time of DeepWalk, and the result of node2vec is similar to DeepWalk as they are both random-walk based methods.
From Fig. \ref{fig:8}, we observe that COSINE accelerates the training process of LINE and DeepWalk on three datasets by the reduction of training samples and the parameters sharing.

Besides the running time of compressed models, we examine the overhead of graph partitioning and group mapping.
In Fig. \ref{fig:12}, we report the time of overhead and two compressed models on three datasets. After preprocessing, all model can reuse the group mapping result.
So, the influence of overhead will reduce when more models reuse the result.
We observed that the influence of preprocessing is much small on Youtube and Flickr.
In Yelp, the overhead is close to LINE's training time and is very small compared to COSINE-DW's time.
When we add overhead to COSINE-LINE's time on Yelp, we found the total time is also reduced by 70\%, which shows the time efficiency of COSINE.
To sum up, the preprocessing, graph partitioning and group mapping, can significantly and consistently reduce the running time of baseline and the overhead of preprocessing has little influence on the total time.

\subsection{Different Partitioning Algorithms}

In this subsection, we examined three candidate partitioning algorithms and discuss the influence of different algorithms. They are:

\textbf{K\scriptsize{A}\small FFP\scriptsize{A}\small ~\cite{sanders2011engineering}} is a multilevel graph partitioning framework which contributes a number of improvements to the multilevel scheme which lead to enhanced partitioning quality. This includes flow-based methods, improved local search and repeated runs similar to the approaches used in multigrid solvers.

\textbf{ParHIP~\cite{meyerhenke2017parallel}} adapts the label propagation technique for graph clustering. By introducing size constraints, label propagation becomes applicable for both the coarsening and the refinement phase of multilevel graph partitioning.

\textbf{mt-metis~\cite{lasalle2013multi}} is the parallel version of METIS that not only improves the partitioning time but also uses significantly less memory.

ParHIP is implemented with MPI library, which means it can be processed in parallel. But, the values of partitions number in the COSINE framework is much bigger than regular values.
So, it causes more communication between different processes and makes partitioning slower. Through experiments, we found the single process is the best setting for ParHIP for time efficiency.

In contrast to ParHIP, mt-metis is implemented with OpenMP library, which means it can use several threads in parallel and share memory between threads. For time efficiency, we use eight threads in experiments.

\begin{figure}[!htb]
\centering
\subfigure[Multi-label classification]{                  
\begin{minipage}{0.45\linewidth}
\centering
\includegraphics[scale=0.23, trim={3cm 7cm 0 5cm}]{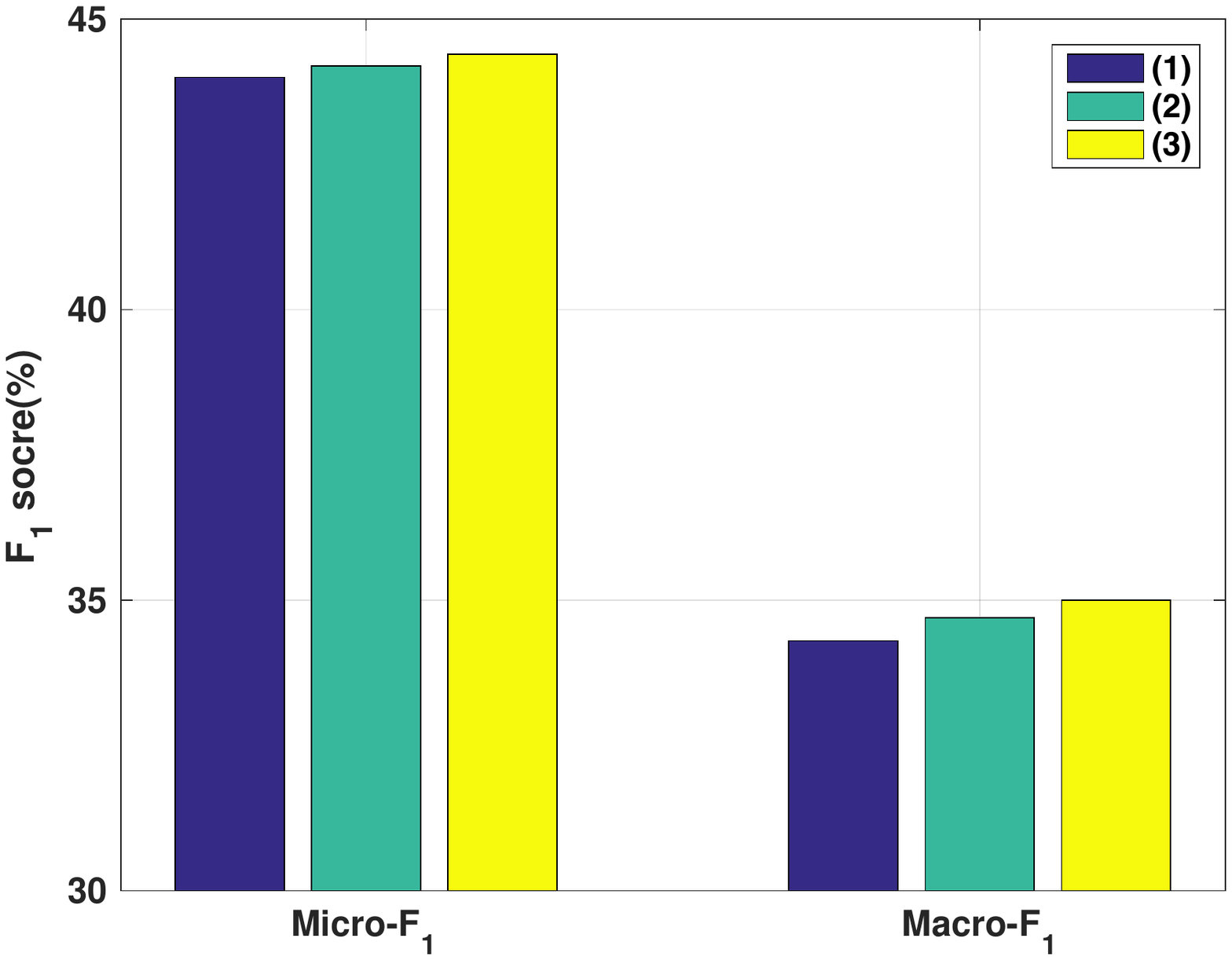}
\label{fig:7:a}         
\end{minipage}}
\subfigure[Link prediction]{                    
\begin{minipage}{0.45\linewidth}
\centering
\includegraphics[scale=0.23, trim={2cm 7cm 0 5cm}]{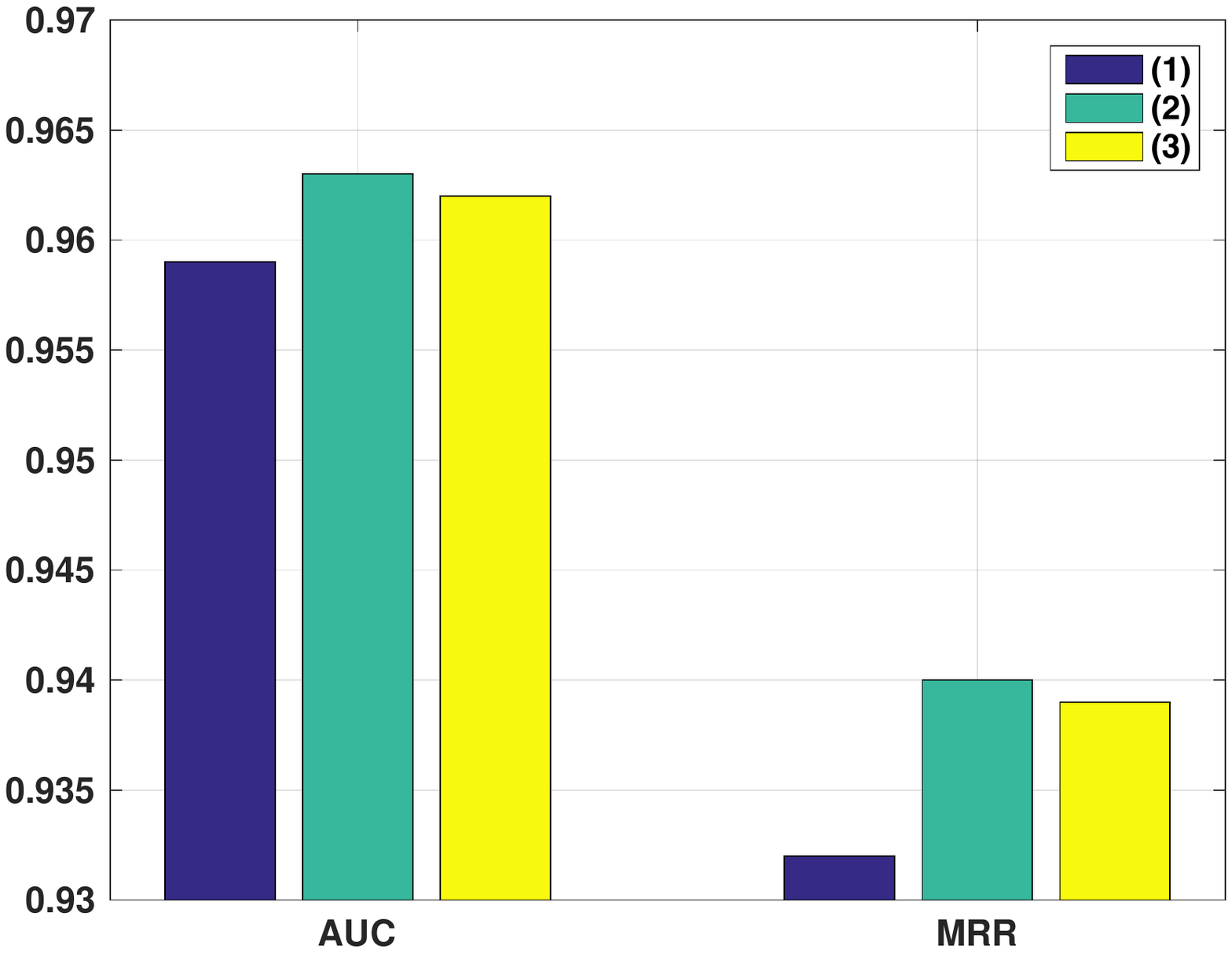}
\label{fig:7:b}             
\end{minipage}}
\caption{Performances of classification and link prediction on Youtube network w.r.t. graph partitioning algorithms: (1) KaFFPa, (2) ParHIP, and (3) mt-metis.}
\label{fig:7}                                                     
\end{figure}

Firstly, we examine the performance of COSINE in Youtube dataset with different partitioning algorithms. We select LINE$_{2nd}$ as baseline model and report the $F_1$ scores of classification, link prediction results using Dot product in Fig. \ref{fig:7}. We observed that three algorithms have similar performance in classification tasks while ParHIP and mt-metis outperform KaFFPa in link prediction. In general, there are tiny differences for performances of algorithms.

\begin{figure}[!htb]
\centering
\includegraphics[scale=0.4, trim={0 7cm 0 7cm}]{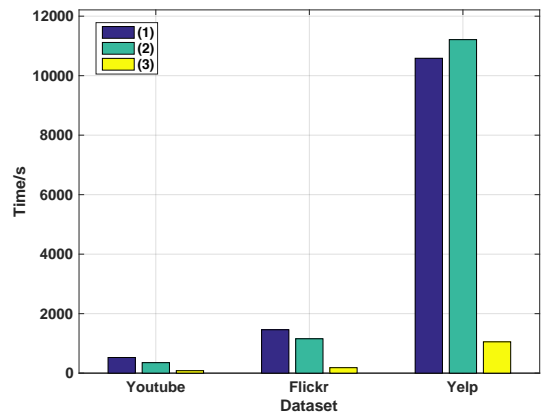}
\caption{Graph partitioning times w.r.t. algorithms and datasets: (1) KaFFPa, (2) ParHIP, and (3) mt-metis.}
\label{fig:6}                                                     
\end{figure}

Secondly, we examine the time cost of different partitioning algorithms. We run these algorithms on three large-scale social networks and record the graph partitioning time respectively. Results are shown in Fig. \ref{fig:6}. We see that the times of ParHIP and KaFFPa are close; ParHIP is faster in Youtube and Flickr, and KaFFPa is faster in Yelp. The mt-metis with multi-threads just takes nearly 10\% of KaFFPa's time to complete partitioning tasks.

In conclusion, we select mt-metis as the partitioning algorithms in the COSINE framework for its time efficiency. It takes less time and gives us a competitive performance.

\subsection{Parameter Sensitivity}

The COSINE framework involves a number of parameters, and we examine how the different choices of parameters affect the performance of COSINE. When graph partitioning and group mapping, we need to set walk per vertex $\gamma$, walk length $k$ and the number of groups for each node $n$. 

\begin{figure}[!htb]
\centering
\subfigure[]{                  
\begin{minipage}{0.45\linewidth}
\centering
\includegraphics[scale=0.23, trim={3cm 7cm 0 5cm}]{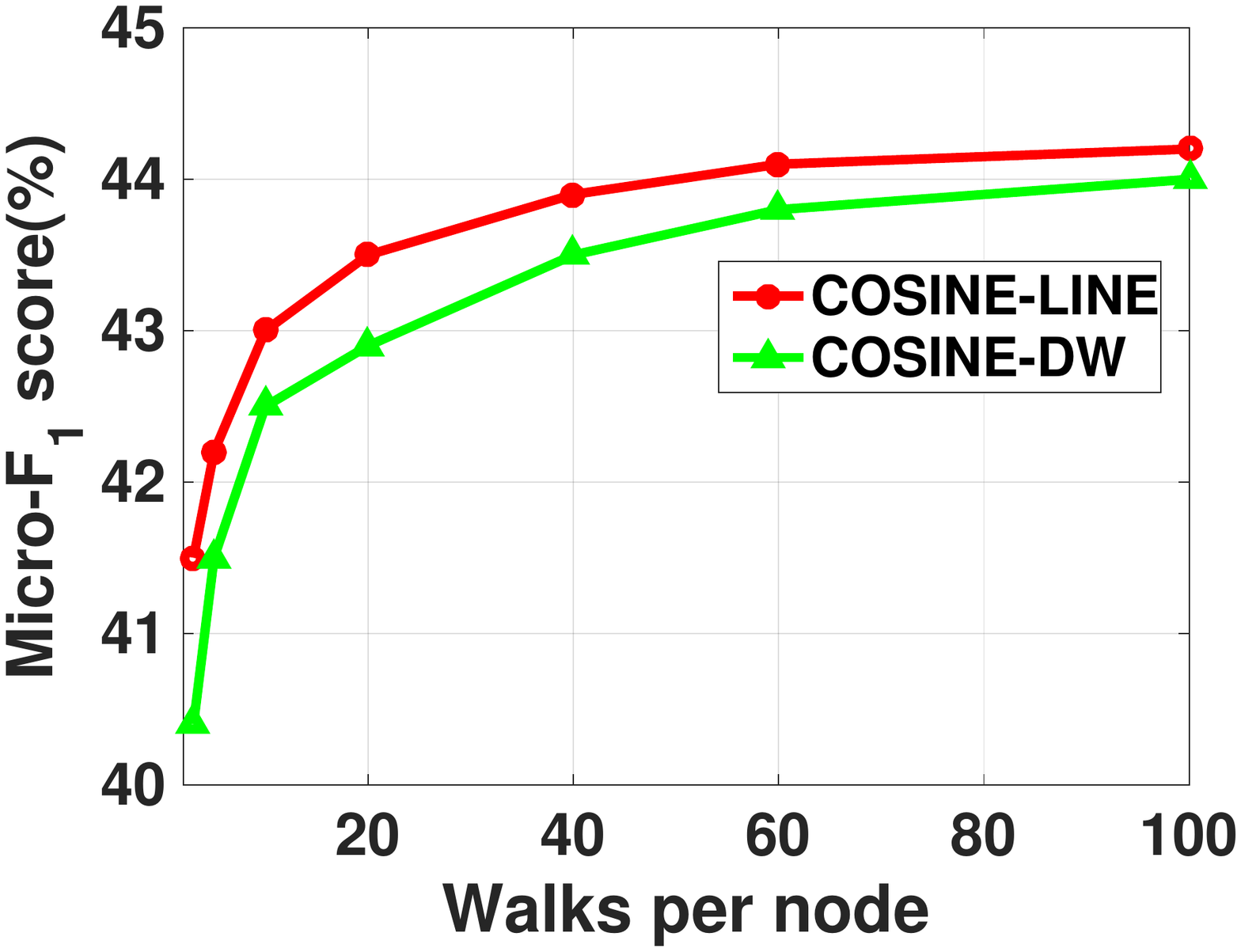}
\label{fig:11:a}         
\end{minipage}}
\subfigure[]{                    
\begin{minipage}{0.45\linewidth}
\centering
\includegraphics[scale=0.23, trim={2cm 7cm 0 5cm}]{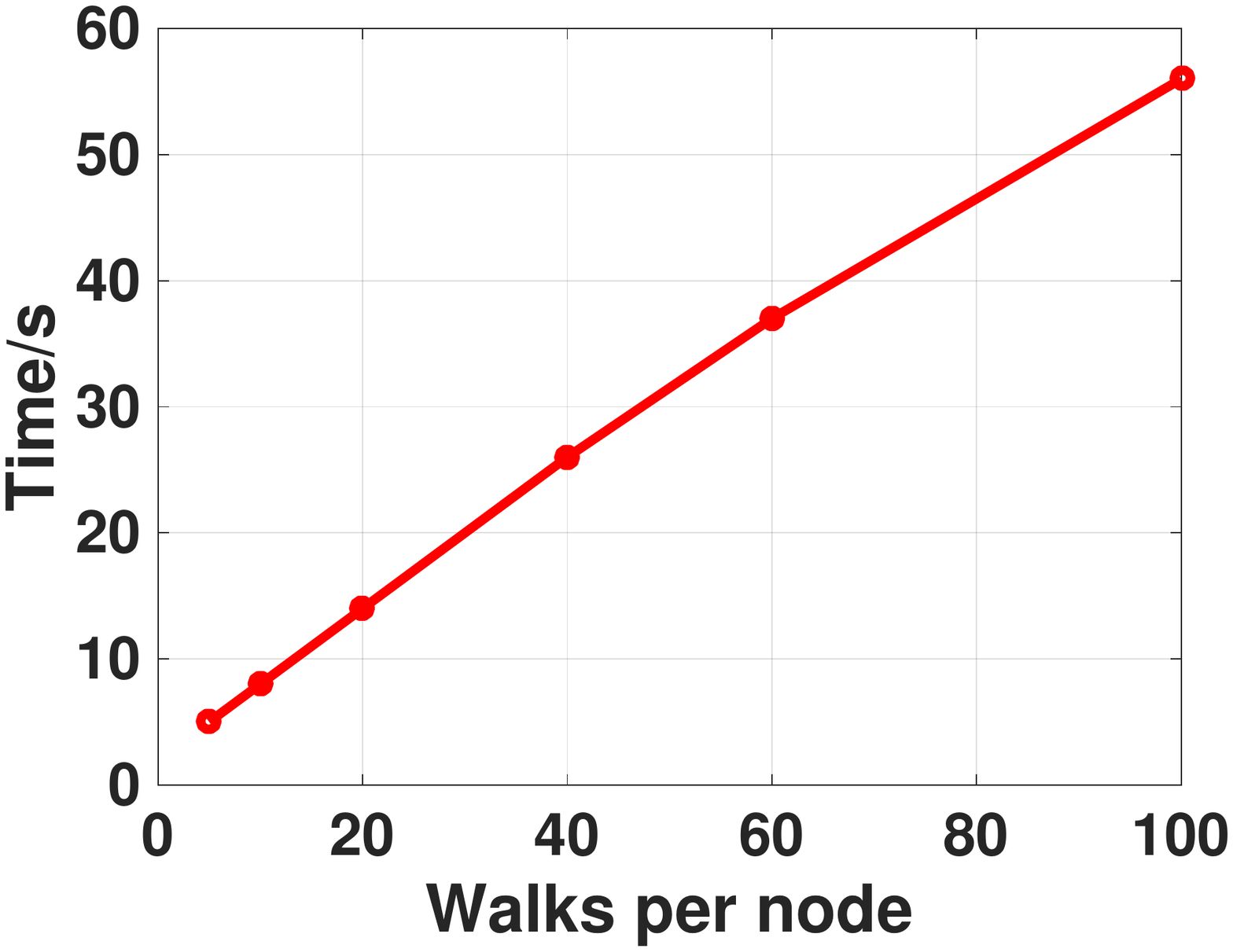}
\label{fig:11:b}             
\end{minipage}}
\caption{(a) Multi-label classification results on Youtube w.r.t walk number per vertex, (b) Group mapping times on Youtube w.r.t walk number per vertex.}
\label{fig:11}                                                     
\end{figure}

Firstly, we select COSINE-LINE and COSINE-DW and evaluate them on Youtube classification task for parameter $\gamma$. In Fig. \ref{fig:11:a}, we show the Macro-F$_1$ scores as a function of parameter $\gamma$. The performance of COSINE improves as walk per vertex $\gamma$ increases. This increase in performance can be explained for more samples from nodes' neighborhood. With a large number of neighbor samples, COSINE understands nodes' local structure better and build a more suitable group set for each node. Moreover, the performance increase becomes slow when the value of parameter $\gamma$ is bigger than 60. It means adding more neighbor samples just offer little local structure information. For time efficiency, we set $\gamma = 100$ for our main experiments.

We implement a multi-threads group mapping program. In Fig. \ref{fig:11:b}, we show the time cost as a function of parameter $\gamma$ with eight threads. We observed that there is a linear relation between time cost and walk per vertex $\gamma$, which is consistent with our algorithm design. And group mapping with 100 walks per node just needs 55 seconds for a network with one million nodes, which shows that group mapping is very scalable.

Secondly, we select COSINE-LINE and evaluate node classification on Youtube dataset and show how walk length $k$ and the number of groups for each node $n$ influence the performance. As we keep the memory usage the same, we need to change the total number of partition groups when we change the number of groups for each node, $n$. For example, the total number of partition groups will decrease if the number of groups for each node, $n$, increases.

\begin{figure}[!htb]
\centering
\includegraphics[scale=0.4, trim={0 7cm 0 7cm}]{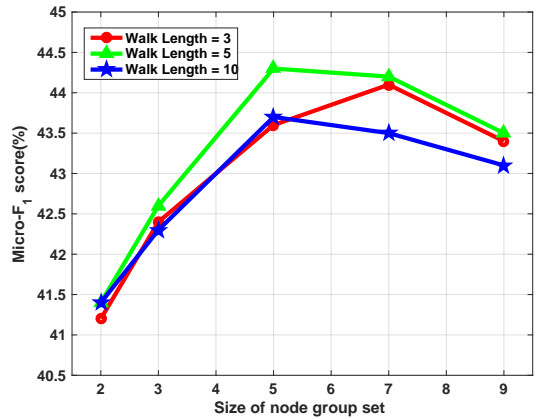}
\caption{Multi-label classification results on Youtube w.r.t walk length and size of nodes' group sets.}
\label{fig:12}                                                     
\end{figure}

In Fig. \ref{fig:12}, we show Micro F$_1$ scores as a function of the size of node group set $n$ with different walk length $k$. From this figure, we observe that:

\begin{enumerate}[(1)]
\item The best value of walk length $k$ is 5. The small value cannot catch the high order proximity in the network while the big value will introduce more noise to the node group set due to the visit of some unrelated node.
\item When setting walk length $k=5$, the best value of groups number for each node $n$ is 5. The small value cannot ensure an effective parameter sharing while the big value causes redundancy in group sets as the neighbor nodes' set has lots of repeated elements.
\end{enumerate}

In summary, group mapping should help model share parameters effectively and efficiently. Each group should belong to more nodes for parameters sharing and the intersection between groups' nodes should be small to reduce the redundancy. We set $n=5, k=5$ in our experiments for effectiveness and efficiency.

\section{Conclusion and Future Work}

In this work, we proposed to leverage graph partitioning to parameter sharing on network embedding, which reduces the redundancy of \textit{embedding lookup} methods.
Our approach is capable of learning compressive high-quality large-scale network embedding in limited memory and taking full use of training samples for accelerating the learning process.
Analyses of the state-of-art network embedding methods with COSINE framework show that our approach significantly improves the quality of embeddings and reduces the running time.

In the future, we will explore the following directions:

\begin{enumerate}[(1)]
\item In this work, we apply three different algorithms for graph partitioning, which shows that all of the algorithms provide competitive results compared to uncompressed methods.
So, we aim to design a more efficient partitioning algorithm for parameters sharing in network embedding.
\item We seek to investigate the extensibility of our model on heterogeneous information network (HIN). In HIN, we need to adjust the group mapping algorithm for incorporating the heterogeneous information comprehensively.
\end{enumerate}

%



\ifCLASSOPTIONcaptionsoff
  \newpage
\fi



%

\bibliographystyle{IEEEtran}
\bibliography{citation}

\end{document}